\documentclass[%
reprint,
superscriptaddress, 
amsmath,amssymb,
aps,
prb,
prstab,
floatfix,
longbibliography
]{revtex4-2}

\usepackage[caption=false]{subfig}
\usepackage{multirow}
\usepackage{graphicx}
\usepackage{dcolumn}
\usepackage{bm}
\usepackage{threeparttable}
\usepackage{color}
\usepackage[dvipsnames]{xcolor}
\usepackage{chemformula}
\usepackage{xspace}
\usepackage{orcidlink}
\usepackage{amsmath}
\usepackage{tabularx}
\usepackage[normalem]{ulem}

\usepackage{amssymb}
\usepackage[capitalize]{cleveref}
\usepackage{booktabs}

\newcommand{\nips}{\ch{NiPS3}\xspace}
\newcommand{\feps}{\ch{FePS3}\xspace}
\newcommand{\mps}{MPS$_3$\xspace}
\newcommand{\cops}{\ch{CoPS3}\xspace}
\newcommand{\mnps}{\ch{MnPS3}\xspace}

\newcommand{\TN}{$T_\mathrm{N}$\xspace}
\newcommand{\TNexp}{$T_\mathrm{N,exp}$\xspace}

\definecolor{AMK}{RGB}{220, 10, 10}
\definecolor{new}{RGB}{0, 0, 0}
\definecolor{Old}{RGB}{150, 150, 150}

\bibpunct{[}{]}{;}{n}{}{}

\hypersetup{colorlinks=true, linkcolor=blue, citecolor=blue, urlcolor=blue, pdftitle = {{L}aser-induced {D}emagnetization in van der {W}aals XY- and {I}sing-like {A}ntiferromagnets nips3 and feps3}, pdfauthor = {D.V.~Kuntu}}

\begin{document}
\preprint{APS/123-QED}

\title{{L}aser-induced {D}emagnetization in van der {W}aals \\$XY$- and {I}sing-like {A}ntiferromagnets {N}i{PS}$_3$ and {F}e{PS}$_3$}

\author{D.~V.~Kuntu\,\orcidlink{0000-0003-0372-8331}}
\email{dariya.kuntu@mail.ioffe.ru}
    \affiliation{Ioffe Institute, 194021 St.\,Petersburg, Russia}
     
\author{E.~A.~Arkhipova\,\orcidlink{0009-0002-7939-9749}}
     \affiliation{Ioffe Institute, 194021 St.\,Petersburg, Russia}

\author{L.~A.~Shelukhin\,\orcidlink{0000-0001-8538-3773}}
     \affiliation{Ioffe Institute, 194021 St.\,Petersburg, Russia}

\author{F.~Mertens\,\orcidlink{0000-0002-7877-0107}}
     \affiliation{Department of Physics, TU Dortmund University, Otto-Hahn-Straße 4, Dortmund 44227, Germany}
     
\author{M.~A.~Prosnikov\,\orcidlink{0000-0002-7107-570X}}
     \affiliation{Ioffe Institute, 194021 St.\,Petersburg, Russia}

\author{I.~A.~Eliseyev\,\orcidlink{0000-0001-9980-6191}}
     \affiliation{Ioffe Institute, 194021 St.\,Petersburg, Russia}

\author{A.~N.~Smirnov\,\orcidlink{0000-0001-9709-5138}}
     \affiliation{Ioffe Institute, 194021 St.\,Petersburg, Russia}

\author{V.~Yu.~Davydov\,\orcidlink{0000-0002-5255-9530}}
\affiliation{Ioffe Institute, 194021 St.\,Petersburg, Russia}

\author{S.~Ma{\~n}as-Valero\,\orcidlink{0000-0001-6319-9238}}    \affiliation{Instituto de Ciencia Molecular (ICMol), Universidad de Valencia, Paterna, 46980, Spain}

\author{E.~Coronado\,\orcidlink{0000-0002-1848-8791}}
\affiliation{Instituto de Ciencia Molecular (ICMol), Universidad de Valencia, Paterna, 46980, Spain}

\author{M.~Cinchetti\,\orcidlink{0000-0003-0735-8921}}
\affiliation{Department of Physics, TU Dortmund University, Otto-Hahn-Straße 4, Dortmund 44227, Germany}

\author{A.~M.~Kalashnikova\,\orcidlink{0000-0001-5635-6186}}
\affiliation{Ioffe Institute, 194021 St.\,Petersburg, Russia}

\date{\today}

\begin{abstract}
The critical behaviour of laser-induced changes in magnetic ordering is studied experimentally in two-dimensional zigzag antiferromagnets  \textit{XY}-like \nips and Ising-like \feps. 
To examine laser-induced dynamics in flakes of these compounds, we employ time-resolved exchange linear dichroism effect sensitive to zigzag magnetic ordering and independent of the orientation of the antiferromagnetic vector.
In both compounds laser excitation in the vicinity of the absorption edge induces partial quenching of the antiferromagnetic ordering manifested by exchange linear dichroism reduction. 
The amplitude of the effect varies with temperature as the derivative of the antiferromagnetic vector and exhibits a critical behaviour with the exponents corresponding to \textit{XY}- and Ising-models for \nips and \feps, respectively.
Critical slowing down of the demagnetization in the vicinity of N\'eel temperature is found, however, only in \feps.
In contrast, the increase of the demagnetization time  near the ordering temperature in \nips is minor.
We show that the difference in the demagnetization times correlates well with the spin specific heat in both compounds.
Beyond the range of slowing down, the demagnetization times in \nips and \feps are comparable, about 5~--~10~ps, and are 
longer than those reported earlier for \cops and considerably shorter than for \mnps. This points to the importance of the unquenched angular momentum of transition-metal ions in laser-induced demagnetization process.
\end{abstract}

\maketitle

\section{Introduction}
\label{sec:intro}

The phenomenon of ultrafast demagnetization, namely quenching of the magnetic ordering in magnets following excitation with femtosecond laser pulses, has become a challenging topic for experimental and theoretical studies.
It yielded discoveries of single- and multi-shot all-optical magnetic reversal~\cite{Stanciu-PRL2007,Ostler2012,Medapalli2017}, generation of ultrashort spin currents~\cite{Huisman2016} and emission of broadband THz pulses~\cite{Seifert2016}.
In metals, demagnetization is governed by electron-phonon coupling~\cite{Koopmans2010}, leading to characteristic timescales of the process of less than 100 femtoseconds~\cite{Beaurepaire1996,Stamm2007,Carpene2008}.
In insulators, this timescale is defined mainly by spin-lattice coupling, and the demagnetization process develops in a subnanosecond range~\cite{Kimel-PRL2002,Maehrlein-SciAdv2018}. 
The ultrafast demagnetization process is also intrinsically sensitive to critical phenomena occurring in the vicinity of magnetic ordering temperature~\cite{Lopez-PRB2013,Kimling-PRB2014}.
Therefore, the type and dimensionality of the magnetic ordering can play an important role in this process, along with the electronic band structure and coupling of spins to other energy and angular momenta reservoirs.

Layered magnetic materials, such as van der Waals materials~\cite{Susner2017,Wang2018}, organic-inorganic hybrids~\cite{Bellitto2015}, MAX phases~\cite{Ingason2016}, etc., offer an appealing possibility to investigate ultrafast magnetic dynamics in connection to details of magnetic ordering.
Indeed, while having similar crystallographic structure within a family, these materials demonstrate a broad variety of spin arrangements. 
Another common feature of such compounds is a weak interlayer exchange coupling, making them quasi two-dimensional (2D) structures.
As a result, these materials are used for examining the Mermin-Wagner theorem and ways to overcome restrictions on magnetic ordering in the 2D limit~\cite{jenkins_breaking_2022}.
In a few- and single-layer limit readily achievable for, e.g., van der Waals materials~\cite{kuo2016exfoliation, Xu_Liang_Shi_Chen_2013, Mas-Balleste2011, Butler2013}, their magnetic ordering can experience drastic changes such as a transition between different spin alignments~\cite{Huang2017} or quenching of magnetic ordering~\cite{Kim2019}, although the latter remains a matter of further discussion~\cite{Kim_Park_2021}.

Laser-pulse driven spin dynamics in layered magnets has recently become a subject of active studies.
They cover revealing manifestations of a quasi-2D nature of magnetic ordering~\cite{Caretta-PRB2015}, as well as examining coupling between lattice and spins and its effect on laser-driven processes in both subsystems~\cite{Mertens2022,Afanasiev2021,Zhou2022,Khusyainov2023,Lichtenberg-2D2023}. 
Examination of a critical behaviour of the laser-induced spin dynamics takes a central place in many of these studies~\cite{Zhang2021,Khusyainov2023,Lichtenberg-2D2023}, with van der Waals transition metal thiophosphates \mps (M=Mn,Fe,Co,Ni) playing an essential role as the materials offering a variety of antiferromagnetic (AFM) arrangements~\cite{Susner2017}.

In this Article, we report on a comparative study of laser-induced demagnetization in the van der Waals transition metal thiophosphates \nips and \feps.
Within the \mps family, nickel and iron phosphorus trisulfides possess a number of similarities with each other and are very distinct from the other two compounds. 
Namely, \feps and \nips both are zigzag antiferromagnets, and their magnetism is of predominantly spin origin.
Yet, \feps{} demonstrates the 2D~Ising-like critical properties, while those of \nips{} are described by the 2D~$XY$-model. 
By studying the time-resolved dynamics of the exchange linear dichroism (exLD), 
we reveal that the demagnetization degree in both thiophosphates increases in the vicinity of the N\'eel temperature and is governed by the critical behaviour of the order parameter depending on the type of 2D magnetic ordering.
However, only 2D~Ising-like antiferromagnet \feps exhibits a pronounced slowing-down of the demagnetization rate as N\'eel temperature is approached, while in 2D~$XY$-like \nips, such a slowing down in the vicinity of \TN is minor.
Using a two-temperature model for the lattice and the spin responses to the optical excitation, we identify that the difference in demagnetization times stems from distinct spin heat capacity behaviours.

\section{Experimental}
\label{sec:expermental}

\subsection{Materials and samples characterization}

\begin{figure}[!t]
    \centering    
    \includegraphics[width=1\columnwidth]{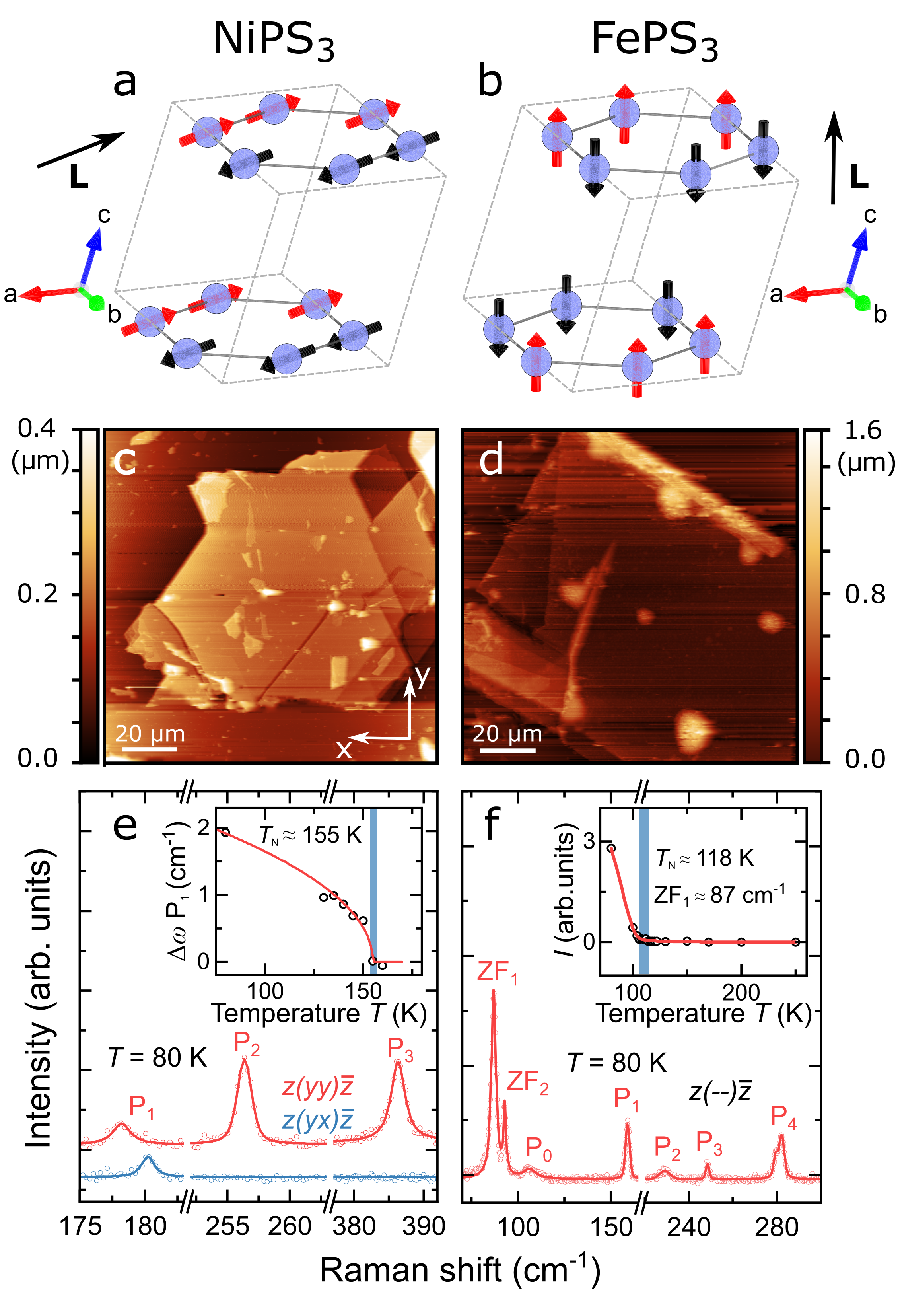}
    \caption{
        \label{fig:exp_scheme}
        (a,b) Magnetic structure of (a) \nips and (b) \feps, blue spheres represent metal atoms. $\mathbf{L}$ is the antiferromagnetic vector. 
        The figure was created using the VESTA~\cite{momma_vesta3_threedimensional_2011}. (c,d) Atomic force microscopy images of (c) \nips and (d) \feps flakes on Si/\ch{SiO2} substrate.
        (e) Polarized Raman spectra of \nips measured at $T=80$~K in parallel [$z(yy)\overline{z}$] and crossed [$z(yx)\overline{z}$] scattering configurations. 
        Inset shows temperature dependence of magnetic-order-induced frequency difference $\Delta \omega$ for phonon lines $P_{1}$ in parallel and cross polarization configurations. 
        The red line represents fit by the function $\propto|T-T_\mathrm{N}|^{2\beta}$.
        (f) Non-polarized Raman spectrum of \feps measured at $T=80$~K. 
        Inset shows the integral intensity $I$ of the low-frequency ZF$_1$ phonon peak (87~cm$^{-1}$) as a function of $T$. 
        Red line is a guide for the eye.
        Blue vertical bars in (e,f) mark \TN extracted from Raman lines temperature dependences.
    }
\end{figure}

\nips and \feps from the family of transition metal thiophosphates \mps are layered van der Waals materials~\cite{Susner2017}.
They belong to a monoclinic crystal system, space group C$2/m$~(\#12)~\cite{Ouvrard1985}.
As shown in~\cref{fig:exp_scheme}(a,b), the transition metal ions are arranged in a honeycomb structure within the $ab$-plane~\cite{JoyVasudevan1992, Wildes_PRB2015}.
The band gap of \nips and \feps is 1.7~eV and 1.5~eV, respectively~\cite{Brec1979}, and is defined by the charge-transfer transitions involving states of P and S ions. 
Within a bandgap there are weaker $d\!-\!d$ transitions of \ch{M^2+} ions in the near-infrared range~\cite{Grasso-SSI1986}.

Among \mps, nickel and iron thiophosphates have the most similarities in terms of spin arrangement.
Furthermore, they take an intermediate place in terms of quenching the orbital momentum of a magnetic ion between \mnps (the zero orbital momentum of Mn$^{2+}$) and \cops (the largest orbital momentum) \cite{JoyVasudevan1992}.
Thus, they suit well for the comparative study of laser-induced demagnetization.
The Néel temperatures are \TN = 155~K for \nips and \TN = 118~K for \feps~\cite{JoyVasudevan1992}.
As illustrated in~\cref{fig:exp_scheme}(a), in \nips, magnetic moments form ferromagnetically~(FM) coupled zigzag chains.
The chains are AFM-coupled within the $ab$-plane and FM-coupled in the adjacent layers along the $c$-axis~\cite{Wildes_PRB2015}.

In samples with a thickness of more than 50~nm, the AFM vector \textbf{L} lies along $a$-axis with a small deviation from the $ab$-plane~\cite{Wildes_PRB2015,Kim-AdvMater2023}. 
This magnetic system can be described by the 2D~$XY$ or $XXZ$~model~\cite{Kim2019}.
In \feps, the zigzag ordering in the $ab$-plane is the same, while
the chains are AFM-coupled in adjacent layers~\cref{fig:exp_scheme}(b). 
In contrast to \nips, the magnetic properties of \feps are described by the 2D~Ising~model.
\textbf{L} is perpendicular to the layers~\cite{Lancon2016}.
The magnetic order is preserved at least down to the bilayer in \nips~\cite{Kim2019} and to the monolayer in \feps~\cite{Lee2016,Kim_Park_2021}.

\textcolor{new}{Flakes of \nips and \feps were obtained from the crystals grown by solid state reaction, as detailed in~\cite{Siskins2020, Mertens2022}. 
They were mechanically exfoliated using adhesive tape inside an argon glovebox, and deposited onto Si/285 nm SiO$_2$ substrates.}
Samples were characterized by atomic force microscopy, as shown in~\cref{fig:exp_scheme}(c,d).
The lateral sizes of \nips and \feps flakes are $\approx$~80 and $\approx$~90~$\mu$m, respectively.
Average flake thicknesses amounted to 200~nm and 180~nm respectively, i.e. their properties are expected to correspond to those of the bulk samples.

For a detailed characterization, Raman measurements were performed using a LabRAM HR Evolution UV-VIS-NIR (Horiba, France) spectrometer equipped with a confocal microscope.
The spectra were measured using continuous-wave excitation at the 532 nm (2.33 eV) line of a Nd:YAG laser, and the power of the incident light on the sample was limited to 0.4 mW.
The spectra were recorded using an 1800 lines/mm grating and a nitrogen-cooled charge-coupled device detector, while a Leica PL FLUOTAR 50× (NA = 0.55) long working-distance objective lens was used to focus the incident beam into a spot of ~2 µm diameter.
For both samples, the sets of phonon lines in the obtained spectra shown in~\cref{fig:exp_scheme}(e,f) are in good agreement with literature data \cite{Lee2016,Kim2019,Liu-PRL2021}. The temperature dependence of the intensity and frequency of the selected lines is used to examine the transition to the AFM phase. In order to control the sample temperature, a Linkam THMS600 (Linkam Scientific Instruments, UK) temperature control system was used.
Polarized Raman spectra of \nips were measured in parallel $[z(yy)\overline{z}]$ and crossed $[z(yx)\overline{z}]$ scattering configurations, with $z$ being the normal of the flake and $x$ being along its horizontal side, as shown in~\cref{fig:exp_scheme}(c). 
\cref{fig:exp_scheme}(e) shows the spectra obtained at $T=80$~K. 
In the paramagnetic phase of \nips, the phonon peak P$_1$ is observed at a frequency of 178.3~cm$^{-1}$ for both polarizations. 
In the AFM state, the spectral positions of P$_1$ in parallel and crossed polarizations do not coincide due to the splitting of degenerate $E-$like phonon modes, which dramatically increases upon magnetic ordering~\cite{Kim2019}. 
The difference $\Delta \omega$ of P$_1$ frequencies for two polarizations as a function of $T$ is shown in the inset of \cref{fig:exp_scheme}(e). 
Following~\cite{Kim2019}, the temperature dependence is well described by $\Delta\omega\propto|T-T_\mathrm{N}|^{2\beta}$, with a critical exponent $\beta=0.24\pm0.04$ and $T_\mathrm{N}=155\pm0.5$~K, confirming 2D~$XY$ ($\beta=0.231$~\cite{Vaz2008}) ordering and bulk N\'eel temperature, respectively.

The non-polarized Raman spectrum for \feps at $T=80$~K is shown in~\cref{fig:exp_scheme}(f). 
In the paramagnetic phase, there is a broad asymmetric phonon peak at 98~cm$^{-1}$~\cite{Wang2016}. 
Below $T_\mathrm{N}$, two distinct peaks at 87 and 93~cm$^{-1}$  appear due to the zone-folding~(ZF) effect caused by the spin ordering~\cite{Lee2016}.
The intensity of the ZF$_1$ peak at 87 cm$^{-1}$ grows with the decrease of $T$ below \TN.
As one can see from the inset of~\cref{fig:exp_scheme}(f), the integral intensity $I$ of this peak drops to zero at \TN = 118 K, which indicates the transition of \feps to the paramagnetic phase. The obtained \TN value is in good agreement with previous studies of bulk~\feps~\cite{Lee2016,Wang2016}.

\subsection{Pump-probe experiment}
Ultrafast magnetization dynamics in \nips and \feps were measured with the femtosecond magneto-optical pump-probe technique.
The experimental scheme is shown in~\cref{fig:NiPS3_data_curves}(a).
Probe pulses with a duration of 170~fs and a photon energy $E_{\mathrm{pr}}$=1.2~eV were generated by the Yb:KGW regenerative amplifier PHAROS (Light Conversion, Lithuania) at a repetition rate of 5\,kHz.
Optical parametric amplifier ORPHEUS (Light Conversion, Lithuania) was used as a source of pump pulses with photon energy in a range $E_{\mathrm{p}}=$1.63--1.97~eV and duration $\sigma\approx170$~~fs.
Pump pulses had a fluence of $F=$~\textcolor{new}{2.5}~mJ/cm$^2$ and an angle of incidence of about 5$^\circ$. 
The pump spot diameter at a sample was about \textcolor{new}{35}~$\mu$m.
Probe pulses with a fluence of 7~mJ/cm$^2$ were focused on a sample into a spot with diameter of 11~$\mu$m at a normal incidence.
\textcolor{new}{The optical response was checked to be linear at the given probe pulse fluence, as discussed in \cref{sec:App1}}.
Both pump and probe spots were smaller than the lateral sizes of the flakes, \textcolor{new}{and it was possible to focus on an area without visible defects.}
Although within the laser spots the flakes possess some degree of thickness inhomogeneity, the latter is not expected to affect the magnetic \textcolor{new}{or optical} properties because of the overall large thickness of the flakes \textcolor{new}{and the penetration depth of the pump and probe pulses exceeding it. 
They are found to be 300~nm and 4~$\mu$m, respectively, for \nips, and 800~nm and 7~$\mu$m, respectively, for \feps.} 

The sample temperature was varied in the range $T$=78--295~K with the temperature control system Linkam HFS600 (Linkam Scientific Instruments, UK).
Pump-induced probe polarization $\Delta \phi$ and reflectivity $\Delta R$ were measured using balanced and single photodiode schemes, respectively, as functions of pump-probe time delay $\Delta t$. 
In Ref.~\cite{Zhang2021-Giant_OLD}, it was shown that AFM ordering in \feps gives rise to an exLD, i.e., a difference in absorption and refraction for the light incident along the $c$-axis and polarized along and perpendicularly to the zigzag chains, i.e. along the $a$- and $b$-axes, respectively.
This effect stems from the sensitivity of the optical properties to the spin correlations and can also be seen as a magnetic refraction~\cite{Markovin-JETP1979}, which acquires anisotropy because of distinct exchange coupling along and perpendicular to chains.
As a result, exLD gives rise to the rotation of the polarization of the reflected light in the AFM phase.
Since this effect is not sensitive to the orientation of $\mathbf{L}$, it is expected to be present in \nips and \feps when the probe is incident along the flake normal.
This is in contrast to quadratic magneto-optical Voigt and Kerr effects~\cite{Saidl2017}, which, in the considered geometry, require $\mathbf{L}$ to lie in the sample plane and thus should vanish in the \feps flake.

It was found that both $\Delta \phi(\Delta t)$ and $\Delta R(\Delta t)$ were temperature-independent above \TN in both flakes \textcolor{new}{(see \cref{sec:App1}, \cref{fig:appendix2} for details)}. 
Thus, we used the pump-probe curves $\Delta \phi(\Delta t)$ measured at $T=173$~K$~>T_\mathrm{N}$ for both compounds as reference data, which includes temperature-independent laser-induced \textcolor{new}{optical} response of the \textcolor{new}{flakes,} substrate, strain-induced dichroism, etc. 
Below, all presented data are the result of subtraction of the reference curve from a signal at a given temperature, as illustrated in~\cref{fig:NiPS3_data_curves}(b) for the pump-induced rotation.

\section{Results}
\label{sec:results}

\subsection{Laser-induced dynamics of exLD}
\label{subsec:resultsA}

\begin{figure}[!t]
    \centering
    \includegraphics[width=1\columnwidth]{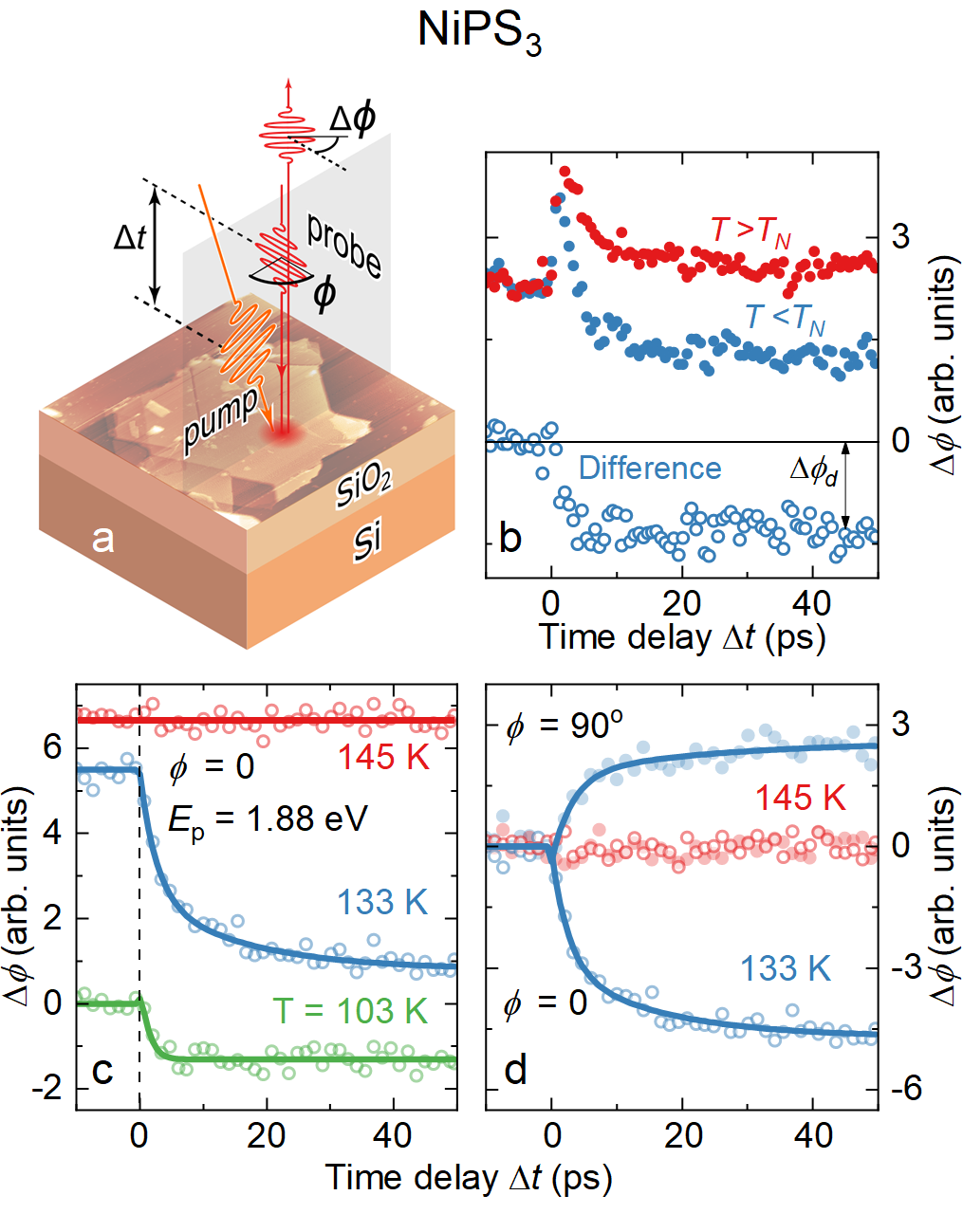}
    \caption{
        \label{fig:NiPS3_data_curves}
        (a) The time-resolved pump-probe experimental scheme.
        Grey shaded area depicts the plane of incidence for the pump pulses. 
        Probe polarization angle $\phi$ is measured from this plane.
        (b) Laser-induced probe polarization rotation $\Delta\phi$ as a function of the delay time $\Delta t$ measured in \nips flake at $T=173$~K~$>$~\TN (red solid symbols) and $T<$~\TN (blue solid symbols) and their difference (open blue symbols). 
        $\Delta\phi_d$ at $\Delta t_d=45$~ps is taken as an amplitude of the laser-induced change of exLD.
        (c) Laser-induced probe polarization rotation $\Delta \phi$ as a function of delay time $\Delta t$ between pump ($E_{\mathrm{p}}=$1.88~eV) and probe pulses measured for different sample temperatures in \nips.
        (d) Transient polarization rotation $\Delta \phi~(\Delta t)$ measured at two probe linear polarizations $\phi=0$ (open symbols) and $\phi=90^{\circ}$ (solid symbols) at $T=$~133~K (blue symbols) and 145~K (red symbols).
        Solid lines are fit according to the~\cref{eq:fit_function_demagnetization}.
    }
\end{figure}

The dynamics of the laser-induced probe polarization rotation in the \nips flake $\Delta\phi$ measured at different sample temperatures and probe polarizations is shown in \cref{fig:NiPS3_data_curves}(c,d).
As seen in~\cref{fig:NiPS3_data_curves}(c), below \TN there is a monotonic change of $\Delta\phi$ at $\Delta t>0$, followed by a saturation.
Relaxation to the initial state takes place at a longer time scale, which is inaccessible in our experiment.
Noticeably, both the magnitude and characteristic time scale of $\Delta\phi(\Delta t)$ evolve with temperature, suggesting that the signal originates from the dynamics of magnetic properties.
Furthermore, below \TN, $\Delta\phi(\Delta t)$ demonstrates pronounced dependence on the incident probe polarization angle $\phi$, with the sign of the laser-induced changes reversed upon rotation of $\phi$ by 90$^\circ$ as shown in~\cref{fig:NiPS3_data_curves}(d).
The laser-induced changes of $\Delta\phi$ are absent for any $\phi$ above \TN.
These temperature and probe polarization dependencies suggest that the measured signal is indeed the transient exLD, and, thus, represents the laser-driven evolution of the AFM ordering in \nips.

\begin{figure}[!t]
    \centering
    \includegraphics[width=1\columnwidth]{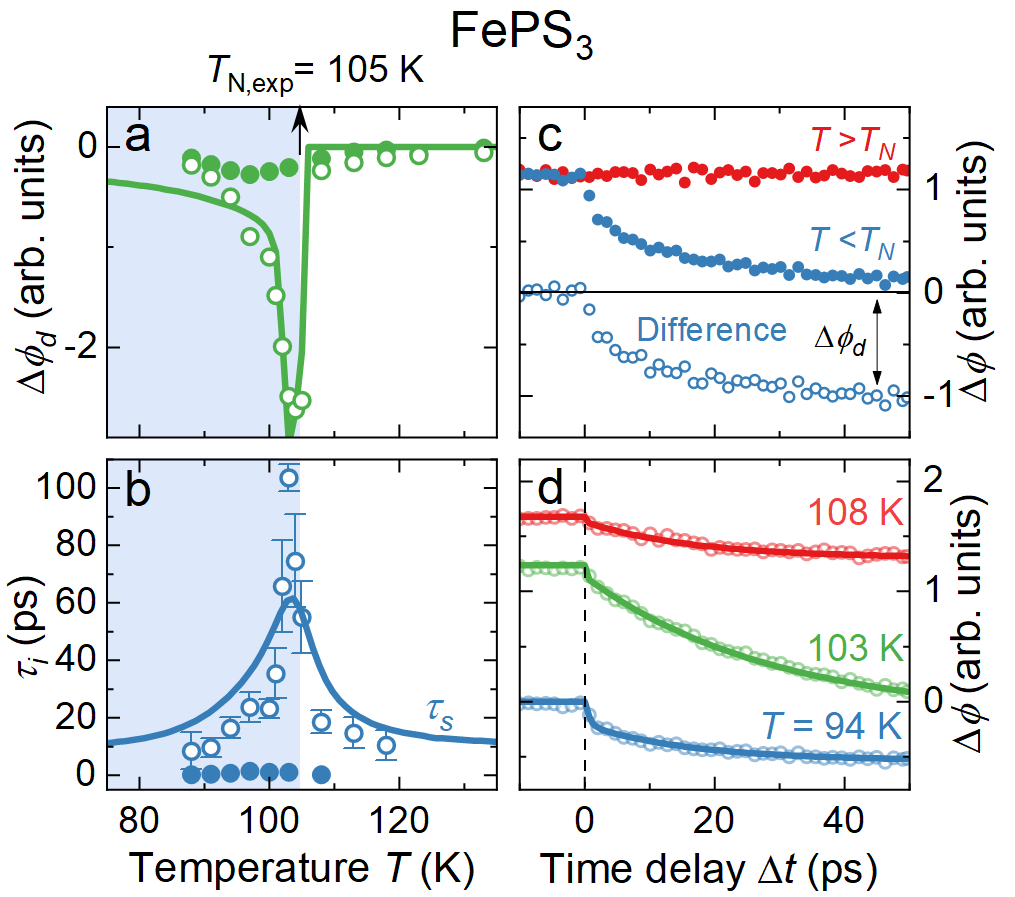}
    \caption{
        \label{fig:FePS3_magnitude_time}
        (a) Transient exLD amplitude $\Delta \phi_d$ (open symbols) and (b) demagnetization time $\tau_2$ (open symbols) in \feps as a function of sample temperature with a fixed $E_\mathrm{p}=$1.97~eV.
        Also shown in (a,b) are the amplitude $\Delta\phi(\Delta t=2$~ps) and characteristic time $\tau_1$ (solid symbols) of the optical contribution to $\Delta\phi$.
        \textcolor{new}{(c) Laser-induced probe polarization rotation $\Delta\phi$($\Delta t$) measured in \feps flake at $T=173$~K~$>$~\TN (red solid symbols) and $T=94$~K~$<$~\TN (blue solid symbols) and their difference (open blue symbols).}        
        (d) Transient polarization rotation~$\Delta \phi$ measured in \feps at different sample temperatures.
        Incident probe polarization angle $\phi=0$.
        \textcolor{new}{Solid (a) green and (b) blue lines are fit according to Eqs.~(\ref{eq:DeltaL_expl};~\ref{eq:2TM}) and~Eqs.~(\ref{eq:DeltaL};~\ref{eq:2TM}), respectively.}
        Solid lines in (d) are fit according to ~\cref{eq:fit_function_demagnetization}. 
    }
\end{figure}

The same measurements were performed for the \feps flake~[\cref{fig:FePS3_magnitude_time}(c)], and yielded qualitatively similar trends, with clearly longer times characterizing the laser-induced change of exLD at some temperatures.
In both samples, \nips and \feps, no dependence of $\Delta\phi(\Delta t)$ 
on pump polarization was observed in the whole studied temperature range.
This points to the essential role of the ultrafast heating by a laser pulse in the measured dynamics.

Taking into account all the experimental observations, we conclude that in both \nips and \feps  below \TN the optical excitation causes the transient partial quenching of the AFM ordering, i. e., demagnetization, induced by the laser heating and manifesting itself via a decrease in exLD.
We note that the laser-induced demagnetization in \feps was earlier demonstrated in~\cite{Zhang2021}, while in \nips no data on this process at the picosecond timescale was reported.

\begin{figure}[!t] 
    \centering
    \includegraphics[width=\columnwidth]{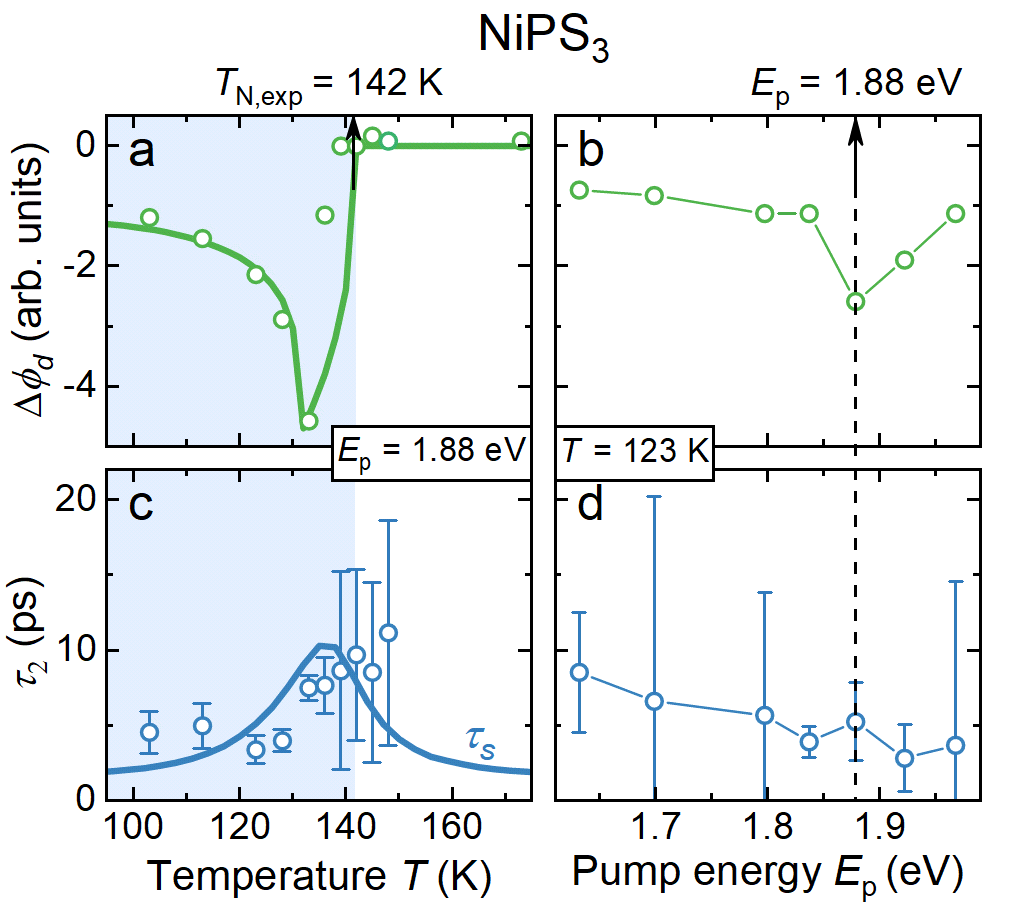}
    \caption{
        \label{fig:NiPS3_magnitude_time_Ep}
        (a,~b) Transient exLD amplitude $\Delta \phi_d$ and (c,~d) demagnetization time $\tau_2$ in \nips as a function of (a,~c) sample temperature $T$ with a fixed $E_{\mathrm{p}}=$1.88~eV and (b,~d) pump photon energy $E_{\mathrm{p}}$ with a fixed sample temperature $T=$~123~K. Incident probe polarization angle $\phi=0$.
        \textcolor{new}{Solid (a) green and (c) blue lines are fit according to Eqs.~(\ref{eq:DeltaL_expl};~\ref{eq:2TM}) and~Eqs.~(\ref{eq:DeltaL};~\ref{eq:2TM}), respectively.} 
    }
\end{figure}

Since the demagnetization is driven by ultrafast laser heating, it can be sensitive to the pump photon energy $E_\mathrm{p}$ due to the spectral dependence of the light absorption coefficient.
To quantify the degree of demagnetization, we used the value $\Delta\phi_d$ at $\Delta t_d=45$~ps [\cref{fig:NiPS3_data_curves}(b)].
At most of the studied temperatures, the transient change of exLD saturates at this time delay in both \nips and \feps.
The study of the laser-induced exLD dynamics in \nips as a function of the pump phonon energy revealed that $\Delta \phi_d$ has a maximum at $E_{\mathrm{p}}=$1.88~eV~[\cref{fig:NiPS3_magnitude_time_Ep}(b)].
This photon energy is in good agreement with the position of an absorption peak reported for \nips~\cite{Banda1986,Joy1992}.
Based on this, we chose such a pump photon energy for measurements of temperature dependences in \nips.
In \feps, no pronounced spectral dependence of the signal was found, and the measurements were performed at $E_{\mathrm{p}}=$1.97~eV.

To analyze the temperature dependence of the laser-induced change of exLD and its features, we examined $\Delta\phi_d(T)$ and also found characteristic times by fitting experimental data at each temperature with a function

\begin{equation}
    \label{eq:fit_function_demagnetization}
    \Delta \phi (\Delta t, T) = \sum\limits_{i = 1}^2 B_i(T) \left[e^{-\Delta t / \tau_i(T)} - 1\right],
\end{equation}

\noindent where $B_i$ are the scaling parameters and $\tau_i$ are the time constants.
Bi-exponential decay was introduced since in \feps one clearly distinguishes faster ($\tau_1$) and slower ($\tau_2$) processes as shown in \cref{fig:FePS3_magnitude_time}(c). 
The physical origin of the bi-exponential decay is discussed further.
We note that the time delay $\Delta t_d=45$~ps at which degree of demagnetization~$\propto\Delta\phi_d$ is analyzed indeed exceeds $3\tau_2$ at most of the studied temperatures in \nips and \feps.

In \nips, one can see that $\Delta\phi_d$ increases sharply
with temperature and then abruptly decreases to zero~[\cref{fig:NiPS3_magnitude_time_Ep}(a)].
This is, again, a characteristic behaviour of the ultrafast laser-induced demagnetization, which is expected to be the largest just below \TN~\cite{Kimel-PRL2002,Maehrlein-SciAdv2018}.
We ascribe the temperature at which the abrupt change of $\Delta\phi_d$ occurs to an experimental N{\'e}el temperature \TNexp, which is below \TN determined from the Raman scattering study (see inset in~\cref{fig:exp_scheme}(e)). 
This is due to \textcolor{new}{unideal thermal contact between the sample and the cold finger in the cryostat, as well as }the excess laser heating not fully dissipating between the excitation events often present in the pump-probe experiments.
The value of \TNexp is discussed below.
The approximation of $\Delta\phi(\Delta t)$ using~\cref{eq:fit_function_demagnetization} gave $B_1=0$.
Thus, $\tau_2$ can be readily ascribed to a characteristic time of the AFM ordering decrease. 
$\tau_2$ grows only slightly in the vicinity of \TNexp and does not exceed 15~ps in the whole studied range, as shown in~\cref{fig:NiPS3_magnitude_time_Ep}(c).

Similarly, in \feps, a sharp increase of $\Delta\phi_d$ is observed as~\TNexp is approached from below, followed by a sharp drop of the signal above it [\cref{fig:FePS3_magnitude_time}(a)].
In contrast to \nips, the approximation of $\Delta \phi(\Delta t)$ using~\cref{eq:fit_function_demagnetization} revealed the bi-exponential decay.
The temperature dependencies of $\tau_1$ and $\tau_2$ are shown in~\cref{fig:FePS3_magnitude_time}(b).
These time constants demonstrate distinct temperature dependences.
$\tau_2$ shows a pronounced increase up to 100~ps in the vicinity of \TNexp, and we ascribe the temperature-dependent $\tau_2$ to the demagnetization time~\cite{Zhang2021}, whereas $\tau_1\approx1.5$~ps is independent of $T$.
Furthermore, the corresponding magnitude of the exLD change, $\Delta\phi(\Delta t=2$~ps), gradually decreases as the temperature is swept across \TNexp towards the paramagnetic phase [\cref{fig:FePS3_magnitude_time}(a)].
We note that the overall laser-induced polarization rotation $\Delta\phi(\Delta t)$ is order of magnitude larger in \feps as compared to \nips.
Thus, the faster contribution to $\Delta\phi(\Delta t)$ in \nips, if present, could be hindered by the sensitivity of the experimental technique.

Recently, fast demagnetization with a characteristic time of 1--5~ps was reported for another crystal from the \mps family \cops ~\cite{Khusyainov2023}.
It was suggested that the unquenched orbital momentum of \ch{Co^2+} ions and strong spin-orbit coupling facilitate a faster demagnetization process in this compound.
However, the degree and characteristic time of demagnetization in \cops demonstrated a pronounced increase near \TN.
Thus, we can rule out such an origin of the fast contribution to $\Delta\phi(\Delta t)$ with $\tau_1$ in \feps, although Fe$^{2+}$ has unquenched orbital momentum as well.
The demagnetization process with two faster and slower stages was also reported for a van der Waals ferromagnet \ch{Fe3GeTe2}~\cite{Lichtenberg-2D2023}.
In contrast to metallic \ch{Fe3GeTe2}, 
spin-flip electron-phonon scattering responsible for the fast demagnetization~\cite{Koopmans2010} is negligible in non-metallic \feps, and the two-stage demagnetization in \feps can be ruled out.

To summarize the experimental findings, ultrafast demagnetization triggered by laser-induced heating is observed by means of the time-resolved exLD in both \nips and \feps.
The demagnetization degree in both compounds possesses a sharp increase as the N\'eel temperature is approached.
Demagnetization times, however, demonstrate distinct behaviour, slowing down in the vicinity of the transition temperature found only in \feps.
The underlying physical picture of the differences in the observed critical behaviour of demagnetization dynamics in \nips and \feps is discussed in the following section.
In \feps, there is an additional faster process leading to changes in exLD with an origin distinct from that responsible for the demagnetization.

\subsection{Numerical simulations of laser-induced dynamics of order parameter}
\label{subsec:Numerical}

ExLD scales as a squared order parameter $\mathbf{L}$ with temperature~\cite{Zhang2021-Giant_OLD,Demokritov-JETP1987}.
According to the Landau phase transition theory~\cite{LANDAU1980446}, the order parameter near \TN varies with temperature following the power law:

\begin{gather}
    \label{eq:powerLaw}
    \left\{
    \begin{split} 
        L \propto \left(T_\mathrm{N}-T\right)^{~\beta}, T<T_\mathrm{N}
        \\
        L=0, T \geqslant T_\mathrm{N}
    \end{split}
    \right.
\end{gather}

\noindent where $\beta$ is the critical exponent. It characterizes the magnetic phase transition and is specific to a model describing the system.
\textcolor{new}{Consequently, an exact expression for the laser-induced change of exLD $\Delta \phi_d$ can be written as}

\begin{equation}
    \label{eq:DeltaL_expl}
    \begin{aligned}
      \textcolor{new}{\Delta\phi_d (T)} & \textcolor{new}{= D (L^2-L_0^2)} \\
        & \textcolor{new}{= D \left[\left(T_\mathrm{N}-(T+\Delta T_s)\right)^{2\beta} - \left(T_\mathrm{N}-T\right)^{2\beta}\right],}
    \end{aligned}
\end{equation}

\noindent \textcolor{new}{where $\Delta T_s$ is the laser-induced change of the effective spin temperature and $D$ is the scaling coefficient. 
In order to show a relation between the dynamics of laser-induced heating of the spin subsystem and $\Delta\phi_d (T,\Delta t)$ explicitly and to separate an effect of the critical behaviour described by $\beta$ from the effect of the spin temperature increase, we introduce the linear approximation of Eq.~(3). 
The latter is valid under the assumption of relatively small laser-induced changes of $L$, i.e., at temperatures $T$ and $T+\Delta T_s$ being far enough from \TNexp.
Then one gets $\Delta\phi_d\propto L\Delta L$, and the expression Eq.~(3) is reduced to a form}

\begin{equation}
    \label{eq:DeltaL}
    \begin{aligned}
        \Delta\phi_d (T) & = D L \frac{dL}{dT}\Delta T_s(T) \\
        & = -2D\beta(T_\mathrm{N}-T)^{2\beta-1}\Delta T_s(T).
    \end{aligned}
\end{equation}

As readily seen from~\cref{eq:DeltaL}, demagnetization time $\tau_2$ correlates with the time required for the heating of the spin subsystem $\tau_s$.
The temperature evolution of $\Delta\phi_d$, in turn, depends on both $\beta$ and the heating of the spin system.
To analyze the response of the spins to optical excitation we employ a two-temperature model~\cite{anisimov1974electron}, considering that \textcolor{new}{in laser-driven dielectrics energy and angular momentum transfer to spins is realized from the phonon subsystem [for a review, see Ref.~\cite{Kirilyuk-RMP2010}] and, thus, it is sufficient to consider only these two reservoirs with effective temperatures $T_s$ and $T_p$, given the electron-phonon relaxation is fast and is of order 2~ps (see App.~\ref{sec:App1} for details).} 
Here we assume that the energy delivered by the pump pulse is absorbed by electrons through $d\!-\!d$ and charge-transfer transitions and rapidly deposited to the lattice.

We solve the following system of equations with the initial condition $T_p=T_s=T$:

\begin{gather}
    \label{eq:2TM}
    \left\{\begin{split}
        \frac{\rho}{\mu}C_p\left(T_p\right)\frac{dT_p}{dt}&=g_{sp}\left(T_s-T_p\right)+P(t), \\
        \frac{\rho}{\mu}C_s\left(T_s\right)\frac{dT_s}{dt}&=g_{sp}\left(T_p-T_s\right),
     \end{split}\right.
\end{gather}

\noindent where $C_p$ and $C_s$ are lattice and magnetic molar specific heat, respectively, and $g_{sp}$  is the spin-lattice coupling constant.
\textcolor{new}{$C_p(T)$ was determined by an approximation of experimental data on the total specific heat $C(T)$ reported in~Ref.~\cite{Siskins2020} with the Debye function~\cite{debye_zur_1912}. Magnetic heat capacity was then obtained as $C_s(T)=C(T)-C_p(T)$.}
$\rho$~and~$\mu$ are the density and the molar mass, respectively.
$P(t)$ is the absorbed laser power density found for the flake with a thickness $d\approx200$~nm as~\cite{walowski2012PhD}

\begin{gather}
    \label{eq:Pt}
    P(t)=AF\frac{\left(1-e^{-\alpha d}\right)}{d}\frac{2\sqrt{\ln2}~e^{-4\ln2~t^2/\sigma^2}}{\sigma\sqrt{\pi}},
\end{gather}

\noindent where $A$ and $\alpha$ are absorptance and absorption coefficient, respectively.
\textcolor{new}{Absorptance $A = 1-R$ was calculated using the Fresnel formula for reflection $R$ at the air-\mps interface with a complex refractive index $n$.
Contributions from the reflections at other interfaces do not exceed 10~\% and are not taken into account.}
Other parameters used in calculations are presented in~\cref{tab:Calc_parameters}.

\begin{table}
\caption{
        \label{tab:Calc_parameters}
        Material parameters used in calculations, spin-phonon coupling constant $g_{sp}$ and critical exponent $\beta$ obtained from the fit for \nips and \feps. }    
    \begin{tabularx}{1\columnwidth}{p{3cm}p{2.5cm}p{2.5cm}}
        \toprule
        Parameter                   &   \nips                               &   \feps                                       \\   
        \midrule
        $\rho$~(kg~m$^{-3}$)          &   3200~\cite{Jain2013_MP}             &   2900~\cite{Jain2013_MP}                     \\ 
        $\mu$~(kg~mol$^{-1}$)         &   0.186                               &   0.183                                       \\
        $n$                         &   $3.7+0.5i$~\cite{Piacentini1982}    &   $2.9+0.3i$~\cite{Zhang2022}                 \\
        $\alpha$~(m$^{-1}$)           &   $3.2\cdot10^6$~\cite{Belvin2021}    &   $1.2\cdot10^6$~\cite{Budniak2022, Brec1979} \\
        $g_{sp}$~(Wm$^{-3}$K$^{-1}$)    &   \textcolor{new}{$(3.2\pm0.8)\cdot10^{16~\dagger}$}  &  $(1.2\pm0.4)\cdot10^{16~\dagger}$\\
        $\beta$ & $0.21\pm0.06^\dagger$             &   $0.10\pm0.03^\dagger$                      \\
        \bottomrule
    \end{tabularx}
\textsuperscript{\textdagger}This work
 
\end{table}

Solving~\cref{eq:2TM}, we find both the temporal evolution of the spin temperature $\Delta T_s(\Delta t)$ and the heating of the spin system $\Delta T_s$ at time delay $\Delta t_d$.
\textcolor{new}{In \nips and \feps the increase in the spin temperature $\Delta T_s$ is below 20 and 10~K, respectively, and has a minimum in the vicinity of \TNexp.
This justifies a direct comparison of the characteristic time of the spin temperature increase $\tau_s$ and the demagnetization time $\tau_2$ according to the linearized expression [Eq.~(4)].}
By fitting $\Delta T_s(\Delta t)$ with a function $1-e^{-\Delta t/\tau_s(T)}$, we obtain $\tau_s$ and compare it with the demagnetization time $\tau_2$ at different initial temperatures.
Note that the fitting procedure also involved correction for \textcolor{new}{\TNexp~$<$~\TN due to unideal thermal contact in the cryostat and excess laser heating}.
It was found to be $\approx13$~K in both samples, which yielded \TNexp~$=142$~K for \nips and \TNexp~$=105$~K for \feps.
Since $\tau_s$ is essentially the spin-lattice thermalization time and is determined by the spin-lattice coupling constant $g_{sp}$, this allows us to find values of $g_{sp}$.
Figs.~\ref{fig:NiPS3_magnitude_time_Ep}(c)~and~\ref{fig:FePS3_magnitude_time}(b) show the calculated thermalization time $\tau_s(T)$ plotted along with $\tau_2$ for \nips and \feps, respectively.
Reasonable agreement with the experimentally obtained temperature dependence of the demagnetization time is obtained for both compounds at the values of $g_{sp}$ listed in~\cref{tab:Calc_parameters}.

As a next step, we fitted the temperature dependence of $\Delta \phi_d (T)$ below $T_{N,exp}$ with the function~[\cref{eq:DeltaL_expl}] using scaling coefficient $D$ and critical exponent $\beta$ as parameters [see solid green lines in Figs.~\ref{fig:NiPS3_magnitude_time_Ep}(a)~and~\ref{fig:FePS3_magnitude_time}(a)].
We obtained $\beta=0.21\pm0.06$ for \nips close to the values 0.231 corresponding to the 2D~$XY$~model~\cite{Vaz2008,Jose-PRB1977} and $0.24\pm0.04$ obtained from the temperature-dependent Raman scattering lines [inset in~\cref{fig:exp_scheme}(e)].
For \feps, the critical exponent amounts to $\beta=0.10\pm0.03$ in agreement with the 2D~Ising~model ($\beta=0.125$~\cite{Vaz2008}) and data obtained from M\"ossbauer 
measurements ($\beta=0.16\pm0.01$~\cite{Yao-Dong2004}).

\textcolor{new}{We note that in a range of $\approx10$~K above \TNexp, there is still a weak nonzero transient exLD signal whose time constant $\tau_2$ follows the spin temperature rise time $\tau_s$ [see Figs.~\ref{fig:NiPS3_magnitude_time_Ep}(c) and \ref{fig:FePS3_magnitude_time}(b)].
This signal can originate from the presence of short-range magnetic ordering, which is not accounted for in the model, and, thus, the amplitude of the transient exLD in this range is not described by \cref{eq:DeltaL_expl} [see Figs.~\ref{fig:NiPS3_magnitude_time_Ep}(a) and \ref{fig:FePS3_magnitude_time}(a)].}

\section{Discussion}

The performed analysis accounting for the laser-induced heating and spin-phonon coupling captures the main experimental finding. Transient exLD indeed measures the demagnetization due to laser-induced heating and is governed by the increase in spin temperature.
Because of the thermal nature of the laser-induced changes, transient exLD and demagnetization $\Delta L$ itself possess critical behaviour at \TN defined by the critical exponent $\beta$. 
Criticality agrees well with the one predicted by 2D~$XY$ or 2D~Ising models.

The demagnetization characteristic time is, in turn, defined by the spin-lattice coupling  constant $g_{sp}$ and the temperature dependence of the spin specific heat, $\tau_2=\tau_s=C_s/{g_{sp}}$.
Pronounced slowing down of the demagnetization near the N\'eel temperature observed in \feps is in agreement with previous works. 
In contrast, it is negligible in \nips.
This correlates well with the features of the specific heat in these materials reported earlier in~\cite{Kim_Park_2021,TAKANO2004,Siskins2020}.
The heat capacity of \nips does not have a well-pronounced feature at $T_\mathrm{N}$, while that of \feps demonstrates a pronounced lambda-peak in $C_s(T)$.

We note that the difference in the critical behaviour of demagnetization times in \nips and \feps resembles the recently reported critical behaviour of characteristic times of restoration of the equilibrium magnetic ordering~\cite{Zhou2022}. 
These times are much longer than the demagnetization times and can be associated with the low interplanar thermal conductivity~\cite{Minnich2016} governing the cooling down process. 
In the vicinity of \TN in \feps, this time shows a pronounced slowing down, reaching 1000~ps, and appears to be two orders of magnitude longer than in \nips. 

Spin-phonon coupling constants $g_{sp}$ are found to be of the same order of magnitude for \nips and \feps, see~\cref{tab:Calc_parameters}.
This, in particular, stems from comparable demagnetization times of around 10~ps in these compounds, apart from the temperature range in the vicinity of \TN. 
Note that in~\cite{Zhang2021}, similar demagnetization time was reported for \feps in a bulk limit.
As for the other two \mps compounds, \mnps and \cops, the demagnetization times well below the corresponding \TN were reported to be 30~ps~\cite{Matthiesen2023} and 2~ps~\cite{Khusyainov2023}, respectively.
Adding our data for \nips and \feps to this set reveals that the demagnetization time below \TN in the \mps family decreases in the row {Mn--Fe,~Ni--Co}.
One notes that this trend does not have a direct correlation with the trend in the N\'eel temperatures, since \nips has the highest \TN, \feps and \cops have very close \TN~\cite{Wildes_JPCM2017}, while the \TN for \mnps is as low as 78~K~\cite{Joy1992}.
Also, there is no direct correlation with the magnitudes of magnetic moments in these compounds either~\cite{Joy1992}.
This is in contrast to demagnetization in  metals~\cite{Koopmans2010}.
We could speculate that in \mps, unquenching of orbital momentum, the most present in \cops and absent in \mnps, affects spin-phonon coupling and, thus, demagnetization times.

To address dependences of the demagnetization degree and time on pump photon energy, we note that they demonstrate no features, apart from the local maximum in $\Delta\phi_d$ at $E_\mathrm{p}=1.88$~eV in \nips [\cref{fig:NiPS3_magnitude_time_Ep}(b)]. 
Demagnetization time
\textcolor{new}{does not show any pronounced trend} within the precision of our experiments [\cref{fig:NiPS3_magnitude_time_Ep}(d)].
On the other hand, in the studied spectral range, optical absorption steadily grows with photon energy due to increased contribution from charge-transfer transitions~\cite{Brec1979}, \textcolor{new}{resulting in an increase of the absorbed pump power.}
\textcolor{new}{If the initial temperature is far enough below \TN}, an increase in absorbed laser power $P$ should lead to a slowdown of demagnetization \textcolor{new}{by $\approx20$~\%} with a concomitant increase in its amplitude, as can be shown based on \cref{eq:DeltaL_expl,eq:2TM}.
\textcolor{new}{Such a slowing down cannot be resolved in our experiments [\cref{fig:NiPS3_magnitude_time_Ep}~(c,d)]}.
\textcolor{new}{We note, however, that} the demagnetization degree \textcolor{new}{in \nips appears to be sensitive} to the localized $d\!-\!d$ transition at $E_\mathrm{p}=1.88$~eV in \nips~\cite{Banda1986,Joy1992}\textcolor{new}{. This} may indicate that there can be an efficient channel of energy dissipation from electrons excited to $3d$ sublevels of \textcolor{new}{$\mathrm{Ni}^{2+}$} to phonons involved in spin-phonon coupling $g_{sp}$.
These can be, e.g., shear phonons, whose particular role in demagnetization was revealed recently~\cite{Zhou2022}.
\textcolor{new}{Noticeably, no resonant enhancement is found in \feps, and there are no reports in the literature on any pronounced features in the absorption spectrum of this material in the studied spectral range.
This can be due to a narrower bandgap in \feps as compared to \nips (1.5 vs. 1.7~eV~\cite{Brec1979}), and, thus, a less pronounced contribution to absorption from $d-d$ transitions.
These observations}, however, require further studies.

Finally, we discuss possible reasons for the transient decrease of exLD with the characteristic time $\tau_1$ observed in \feps.
Temperature dependences $\Delta\phi(\Delta t=2~\mathrm{ps})$ and $\tau_1$~[\cref{fig:FePS3_magnitude_time}(a,b)] suggest that the increase in spin temperature and the corresponding decrease of $\mathbf{L}$ cannot be responsible for this signal, as discussed in Sec.~\ref{subsec:resultsA}.
On the other hand, in optical pump-probe measurements, changes in the measured signal may also be caused by the laser-induced changes of the proportionality coefficient \textcolor{new}{$D$} between the polarization rotation and the magnetic characteristic \cite{Koopmans-PRL2000}, that is $\Delta\phi\propto\Delta \textcolor{new}{D} L^2$. 
Such a contribution to $\Delta\phi$ would follow the temperature dependence of static exLD $\phi$ and decrease in the vicinity of \TNexp, as we observed in the experiment [\cref{fig:FePS3_magnitude_time}(a)].
In the case of exLD, \textcolor{new}{$D$} includes optical parameters as well as the exchange coupling $J$.
Measurements of the laser-induced dynamics of reflectivity $\Delta R(\Delta t)$ in \feps revealed that it is characterized by a fast drop with a time constant of 1.5~ps \textcolor{new}{[see \cref{sec:App1} and \cref{fig:appendix2}] which is} close to $\tau_1$.

Thus, it is reasonable to assume that the faster dynamics of exLD is a trivial result of changes in optical properties leading to $\Delta D$.
Alternatively, the faster contribution to $\Delta\phi$ may stem from laser-induced perturbation of $J$ due to rapid energy transfer to phonons, which, in general, can modulate Fe$^{2+}$--S$^{1-}$ bonds responsible for exchange coupling. 
A fast change of $\Delta R$ can serve a measure of this energy transfer rate.
The feasibility of such phonon-modulated exchange and related transient magneto-optical response was investigated in Ref.~\cite{Maehrlein-SciAdv2018} for the case of a dielectric ferrimagnet Y$_3$Fe$_5$O$_{12}$ subjected to THz pump pulses.
We also note that the magnitude of $\Delta R$ was found to gradually increase as the temperature was changed across \TNexp towards lower temperatures, which may point to the sensitivity of the electron-phonon coupling to the magnetic ordering.

\section{Conclusions}

A comparative experimental study of the laser-induced exLD in 2D zigzag antiferromagnets \nips and \feps with \textit{XY}- and Ising-like magnetic ordering revealed that ultrafast demagnetization in these compounds can be well understood in terms of laser-induced heating of the spin system mediated by spin-phonon coupling.
The demagnetization degree shows critical behaviour as N\'eel temperature is approached,  with exponents being in good agreement with 2D \textit{XY}- and 2D Ising-like ordering in \nips and \feps, respectively.
The spin-phonon coupling parameter is found to be close for two compounds.
Similarly, the demagnetization times well below \TN appear to be close and are of the order of 10~ps.
Previously, demagnetization times were reported for all antiferromagnets from the \mps family except for \nips.
Our data complete this row and show that the time decreases with a change of transition metal Mn -- Fe,~Ni -- Co.
This highlights that unquenched orbital momentum, which is the strongest in Co and is absent in Mn, plays an important role in ultrafast demagnetization in \mps.
Apart from similarities, a difference in demagnetization times in \nips and \feps was found in the vicinity of their respective \TN.
The critical slowdown is observed in \feps only and is nearly absent in \nips. 
In agreement with the thermal nature of the laser-induced demagnetization, this difference is found to be related to the different behaviour of spin specific heat in the vicinity of \TN.

\section{Acknowledgements}

We thank R.~M.~Dubrovin for the help with experiments.
We thank D.~Bossini for insightful discussions. 
The work of D.V.K., V.Yu.D. and A.M.K. was supported by Russian Foundation for Basic Research Grant No. 19-52-12065.
M.A.P. acknowledges support of Russian Science Foundation under Grant No. 22-72-00039. 
L.A.Sh. acknowledges Presidential Scholarship Grant No. SP-4623.2022.5.
M.C. and F.M. acknowledge support by the Deutsche Forschungsgemeinschaft through the International Collaborative Research Centre 160 (Projects B9 and Z4). 
E.C. and M.C. acknowledge funding from the European Union’s Horizon 2020 Research and Innovation Programme under Project SINFONIA, Grant No. 964396.
E.C. acknowledges support from the European Union (ERC-2017-ADG-788222), from the Spanish Government (2D-HETEROS PID2020-117152RB-100, co-financed by FEDER, and Excellence Unit “María
de Maeztu” CEX2019-000919-M) and the Generalitat Valenciana (PROMETEO Program and PO FEDER Program, IDIFEDER/2018/061).
E.C. acknowledges the Advanced Materials program, supported by MCIN with funding from European Union NextGenerationEU (PRTR-C17.I1) and by Generalitat Valenciana.

\appendix
\section{\textcolor{new}{Laser-induced reflectivity dynamics}}
\label{sec:App1}
\begin{figure}[t]
\centering
\includegraphics[width=1\columnwidth]{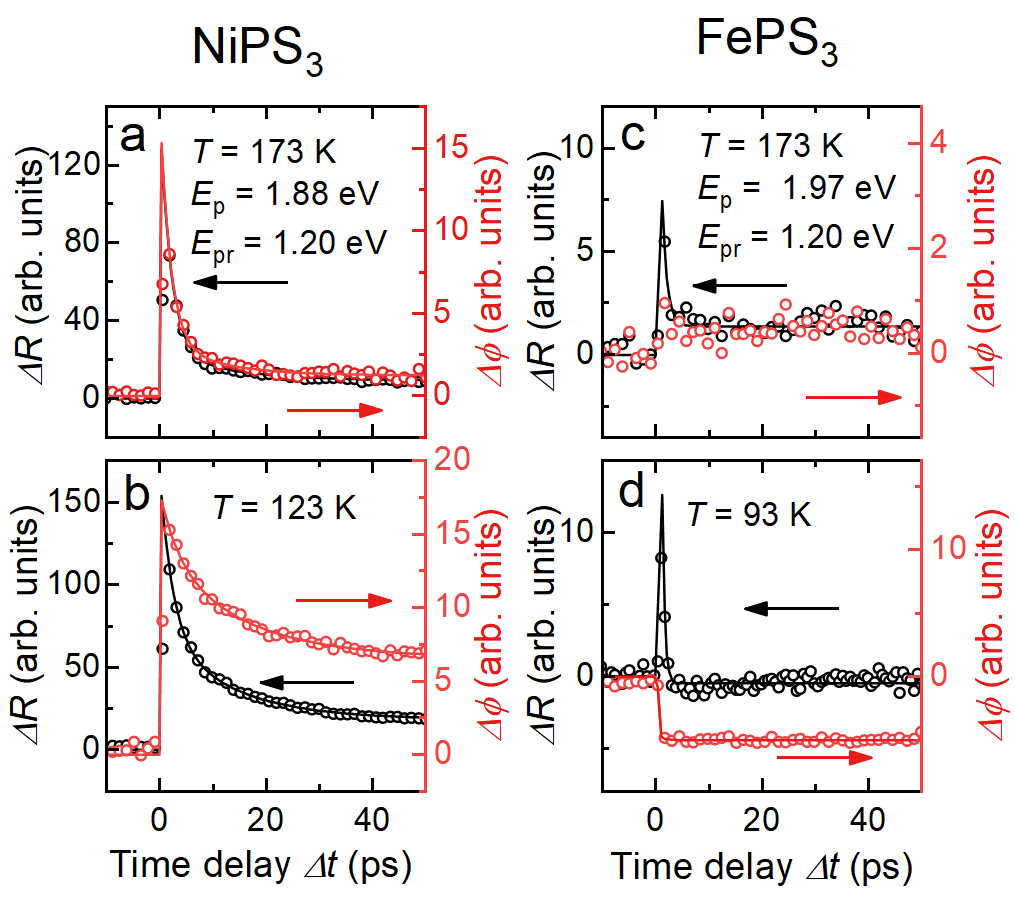}
\caption{\label{fig:appendix2}
\textcolor{new}{Comparison of raw experimental traces for laser-induced probe polarization rotation $\Delta\phi (\Delta t)$ (red) and laser-induced change of reflectivity $\Delta R(\Delta t)$ (black) measured in \nips at (a) $T=173$~K~$>$~\TN and (b) $T=123$~K~$<$~\TN, and in \feps at (c) $T=173$~K~$>$~\TN and (d) $T=93$~K~$<$~\TN.
Solid lines are guides for the eye.
}
}
\end{figure}

\textcolor{new}{In \cref{fig:appendix2}, we present the raw signals $\Delta \phi(\Delta t)$ and $\Delta R(\Delta t)$ in \nips and \feps at temperatures above and below \TN.
The dynamics of the reflectivity change is characterized by the time constant amounting to 2~ps in \nips and 1.5~ps in \feps, which is related to the electron-phonon relaxation time and  depends on temperature only weakly.} 

\textcolor{new}{In both compounds at $T>$~\TN, $\Delta \phi(\Delta t)$ and $\Delta R(\Delta t)$ show similar dynamics [\cref{fig:appendix2}(a,c)].
Thus, we associate transient probe polarization rotation at these temperatures with the changes in the Fresnel coefficients of the flakes, as well as laser-induced changes in the substrate optical constants, strain-induced effects, etc.}

\textcolor{new}{At $T<$~\TN, transients $\Delta \phi(\Delta t)$ and $\Delta R(\Delta t)$ show distinct behaviour [\cref{fig:appendix2}(b,d)].
In particular, probe polarization rotation possesses an additional slower contribution as compared to the dynamics of reflectivity.
This is the contribution from transient exLD, which can be obtained by subtraction of $\Delta \phi(\Delta t)$ measured at $T=$~173~K from the data measured below \TN, as discussed in the main text.
}

\begin{figure}[!h]
\centering
\includegraphics[width=1\columnwidth]{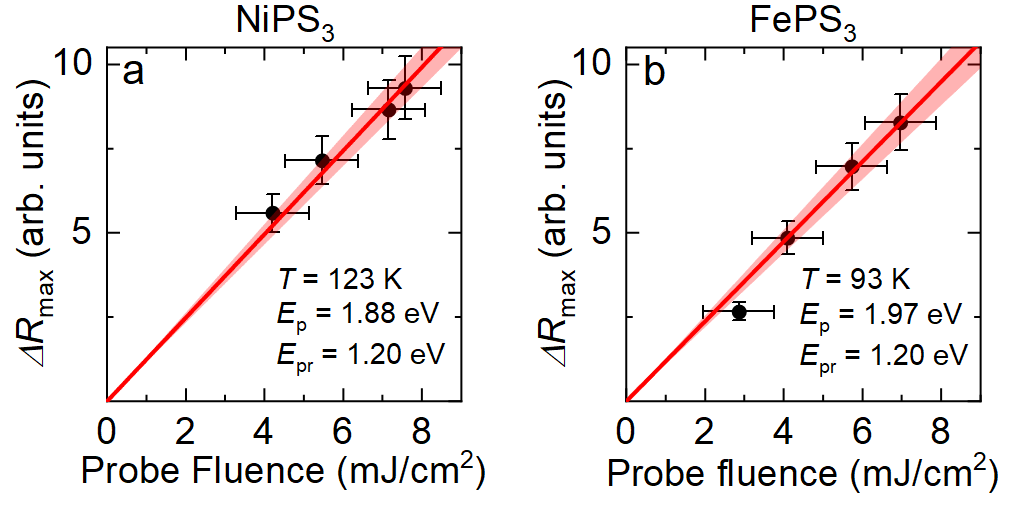}
\caption{
\label{fig:appendix1} 
\textcolor{new}{Magnitude of laser-induced change of reflectivity $\Delta R$ as a function of probe pulse fluence (symbols) for
(a) \nips at $T=123$~K~$<$~\TN, and (b) \feps at $T=93$~K~$<$~\TN. 
The red solid line shows a linear fit with a confidence interval.  }
}
\end{figure}

\textcolor{new}{In order to verify that the measurements of the laser-induced dynamics are performed in the linear regime, we have measured pump-induced changes in reflectivity $\Delta R$ at different probe fluences. 
As shown in Fig.~\ref{fig:appendix1}, there is a linear dependence of $\Delta R$ magnitude measured at $\Delta t = 2$~ps on the probe fluence in both \feps and \nips samples.}

\bibliography{bibliography}

\begin{thebibliography}{68}%
\makeatletter
\providecommand \@ifxundefined [1]{%
 \@ifx{#1\undefined}
}%
\providecommand \@ifnum [1]{%
 \ifnum #1\expandafter \@firstoftwo
 \else \expandafter \@secondoftwo
 \fi
}%
\providecommand \@ifx [1]{%
 \ifx #1\expandafter \@firstoftwo
 \else \expandafter \@secondoftwo
 \fi
}%
\providecommand \natexlab [1]{#1}%
\providecommand \enquote  [1]{``#1''}%
\providecommand \bibnamefont  [1]{#1}%
\providecommand \bibfnamefont [1]{#1}%
\providecommand \citenamefont [1]{#1}%
\providecommand \href@noop [0]{\@secondoftwo}%
\providecommand \href [0]{\begingroup \@sanitize@url \@href}%
\providecommand \@href[1]{\@@startlink{#1}\@@href}%
\providecommand \@@href[1]{\endgroup#1\@@endlink}%
\providecommand \@sanitize@url [0]{\catcode `\\12\catcode `\$12\catcode `\&12\catcode `\#12\catcode `\^12\catcode `\_12\catcode `\%12\relax}%
\providecommand \@@startlink[1]{}%
\providecommand \@@endlink[0]{}%
\providecommand \url  [0]{\begingroup\@sanitize@url \@url }%
\providecommand \@url [1]{\endgroup\@href {#1}{\urlprefix }}%
\providecommand \urlprefix  [0]{URL }%
\providecommand \Eprint [0]{\href }%
\providecommand \doibase [0]{https://doi.org/}%
\providecommand \selectlanguage [0]{\@gobble}%
\providecommand \bibinfo  [0]{\@secondoftwo}%
\providecommand \bibfield  [0]{\@secondoftwo}%
\providecommand \translation [1]{[#1]}%
\providecommand \BibitemOpen [0]{}%
\providecommand \bibitemStop [0]{}%
\providecommand \bibitemNoStop [0]{.\EOS\space}%
\providecommand \EOS [0]{\spacefactor3000\relax}%
\providecommand \BibitemShut  [1]{\csname bibitem#1\endcsname}%
\let\auto@bib@innerbib\@empty
\bibitem [{\citenamefont {Stanciu}\ \emph {et~al.}(2007)\citenamefont {Stanciu}, \citenamefont {Hansteen}, \citenamefont {Kimel}, \citenamefont {Kirilyuk}, \citenamefont {Tsukamoto}, \citenamefont {Itoh},\ and\ \citenamefont {Rasing}}]{Stanciu-PRL2007}%
  \BibitemOpen
  \bibfield  {author} {\bibinfo {author} {\bibfnamefont {C.~D.}\ \bibnamefont {Stanciu}}, \bibinfo {author} {\bibfnamefont {F.}~\bibnamefont {Hansteen}}, \bibinfo {author} {\bibfnamefont {A.~V.}\ \bibnamefont {Kimel}}, \bibinfo {author} {\bibfnamefont {A.}~\bibnamefont {Kirilyuk}}, \bibinfo {author} {\bibfnamefont {A.}~\bibnamefont {Tsukamoto}}, \bibinfo {author} {\bibfnamefont {A.}~\bibnamefont {Itoh}},\ and\ \bibinfo {author} {\bibfnamefont {T.}~\bibnamefont {Rasing}},\ }\bibfield  {title} {\bibinfo {title} {All-optical magnetic recording with circularly polarized light},\ }\href {https://doi.org/10.1103/PhysRevLett.99.047601} {\bibfield  {journal} {\bibinfo  {journal} {Phys. Rev. Lett.}\ }\textbf {\bibinfo {volume} {99}},\ \bibinfo {pages} {047601} (\bibinfo {year} {2007})}\BibitemShut {NoStop}%
\bibitem [{\citenamefont {Ostler}\ \emph {et~al.}(2012)\citenamefont {Ostler}, \citenamefont {Barker}, \citenamefont {Evans}, \citenamefont {Chantrell}, \citenamefont {Atxitia}, \citenamefont {{Chubykalo-Fesenko}}, \citenamefont {El~Moussaoui}, \citenamefont {Le~Guyader}, \citenamefont {Mengotti}, \citenamefont {Heyderman}, \citenamefont {Nolting}, \citenamefont {Tsukamoto}, \citenamefont {Itoh}, \citenamefont {Afanasiev}, \citenamefont {Ivanov}, \citenamefont {Kalashnikova}, \citenamefont {Vahaplar}, \citenamefont {Mentink}, \citenamefont {Kirilyuk}, \citenamefont {Rasing},\ and\ \citenamefont {Kimel}}]{Ostler2012}%
  \BibitemOpen
  \bibfield  {author} {\bibinfo {author} {\bibfnamefont {T.~A.}\ \bibnamefont {Ostler}}, \bibinfo {author} {\bibfnamefont {J.}~\bibnamefont {Barker}}, \bibinfo {author} {\bibfnamefont {R.~F.~L.}\ \bibnamefont {Evans}}, \bibinfo {author} {\bibfnamefont {R.~W.}\ \bibnamefont {Chantrell}}, \bibinfo {author} {\bibfnamefont {U.}~\bibnamefont {Atxitia}}, \bibinfo {author} {\bibfnamefont {O.}~\bibnamefont {{Chubykalo-Fesenko}}}, \bibinfo {author} {\bibfnamefont {S.}~\bibnamefont {El~Moussaoui}}, \bibinfo {author} {\bibfnamefont {L.}~\bibnamefont {Le~Guyader}}, \bibinfo {author} {\bibfnamefont {E.}~\bibnamefont {Mengotti}}, \bibinfo {author} {\bibfnamefont {L.~J.}\ \bibnamefont {Heyderman}}, \bibinfo {author} {\bibfnamefont {F.}~\bibnamefont {Nolting}}, \bibinfo {author} {\bibfnamefont {A.}~\bibnamefont {Tsukamoto}}, \bibinfo {author} {\bibfnamefont {A.}~\bibnamefont {Itoh}}, \bibinfo {author} {\bibfnamefont {D.}~\bibnamefont {Afanasiev}}, \bibinfo {author} {\bibfnamefont {B.~A.}\ \bibnamefont {Ivanov}}, \bibinfo
  {author} {\bibfnamefont {A.~M.}\ \bibnamefont {Kalashnikova}}, \bibinfo {author} {\bibfnamefont {K.}~\bibnamefont {Vahaplar}}, \bibinfo {author} {\bibfnamefont {J.}~\bibnamefont {Mentink}}, \bibinfo {author} {\bibfnamefont {A.}~\bibnamefont {Kirilyuk}}, \bibinfo {author} {\bibfnamefont {T.}~\bibnamefont {Rasing}},\ and\ \bibinfo {author} {\bibfnamefont {A.~V.}\ \bibnamefont {Kimel}},\ }\bibfield  {title} {\bibinfo {title} {Ultrafast heating as a sufficient stimulus for magnetization reversal in a ferrimagnet},\ }\href {https://doi.org/10.1038/ncomms1666} {\bibfield  {journal} {\bibinfo  {journal} {Nature Communications}\ }\textbf {\bibinfo {volume} {3}},\ \bibinfo {pages} {666} (\bibinfo {year} {2012})}\BibitemShut {NoStop}%
\bibitem [{\citenamefont {Medapalli}\ \emph {et~al.}(2017)\citenamefont {Medapalli}, \citenamefont {Afanasiev}, \citenamefont {Kim}, \citenamefont {Quessab}, \citenamefont {Manna}, \citenamefont {Montoya}, \citenamefont {Kirilyuk}, \citenamefont {Rasing}, \citenamefont {Kimel},\ and\ \citenamefont {Fullerton}}]{Medapalli2017}%
  \BibitemOpen
  \bibfield  {author} {\bibinfo {author} {\bibfnamefont {R.}~\bibnamefont {Medapalli}}, \bibinfo {author} {\bibfnamefont {D.}~\bibnamefont {Afanasiev}}, \bibinfo {author} {\bibfnamefont {D.~K.}\ \bibnamefont {Kim}}, \bibinfo {author} {\bibfnamefont {Y.}~\bibnamefont {Quessab}}, \bibinfo {author} {\bibfnamefont {S.}~\bibnamefont {Manna}}, \bibinfo {author} {\bibfnamefont {S.~A.}\ \bibnamefont {Montoya}}, \bibinfo {author} {\bibfnamefont {A.}~\bibnamefont {Kirilyuk}}, \bibinfo {author} {\bibfnamefont {T.}~\bibnamefont {Rasing}}, \bibinfo {author} {\bibfnamefont {A.~V.}\ \bibnamefont {Kimel}},\ and\ \bibinfo {author} {\bibfnamefont {E.~E.}\ \bibnamefont {Fullerton}},\ }\bibfield  {title} {\bibinfo {title} {Multiscale dynamics of helicity-dependent all-optical magnetization reversal in ferromagnetic {C}o/{P}t multilayers},\ }\href {https://doi.org/10.1103/PhysRevB.96.224421} {\bibfield  {journal} {\bibinfo  {journal} {Physical Review B}\ }\textbf {\bibinfo {volume} {96}},\ \bibinfo {pages} {224421} (\bibinfo
  {year} {2017})}\BibitemShut {NoStop}%
\bibitem [{\citenamefont {Huisman}\ \emph {et~al.}(2016)\citenamefont {Huisman}, \citenamefont {Mikhaylovskiy}, \citenamefont {Costa}, \citenamefont {Freimuth}, \citenamefont {Paz}, \citenamefont {Ventura}, \citenamefont {Freitas}, \citenamefont {Bl\"ugel}, \citenamefont {Mokrousov}, \citenamefont {Rasing},\ and\ \citenamefont {Kimel}}]{Huisman2016}%
  \BibitemOpen
  \bibfield  {author} {\bibinfo {author} {\bibfnamefont {T.~J.}\ \bibnamefont {Huisman}}, \bibinfo {author} {\bibfnamefont {R.~V.}\ \bibnamefont {Mikhaylovskiy}}, \bibinfo {author} {\bibfnamefont {J.~D.}\ \bibnamefont {Costa}}, \bibinfo {author} {\bibfnamefont {F.}~\bibnamefont {Freimuth}}, \bibinfo {author} {\bibfnamefont {E.}~\bibnamefont {Paz}}, \bibinfo {author} {\bibfnamefont {J.}~\bibnamefont {Ventura}}, \bibinfo {author} {\bibfnamefont {P.~P.}\ \bibnamefont {Freitas}}, \bibinfo {author} {\bibfnamefont {S.}~\bibnamefont {Bl\"ugel}}, \bibinfo {author} {\bibfnamefont {Y.}~\bibnamefont {Mokrousov}}, \bibinfo {author} {\bibfnamefont {T.}~\bibnamefont {Rasing}},\ and\ \bibinfo {author} {\bibfnamefont {A.~V.}\ \bibnamefont {Kimel}},\ }\bibfield  {title} {\bibinfo {title} {Femtosecond control of electric currents in metallic ferromagnetic heterostructures},\ }\href {https://doi.org/10.1038/nnano.2015.331} {\bibfield  {journal} {\bibinfo  {journal} {Nature Nanotechnology}\ }\textbf {\bibinfo {volume} {11}},\
  \bibinfo {pages} {455} (\bibinfo {year} {2016})}\BibitemShut {NoStop}%
\bibitem [{\citenamefont {Seifert}\ \emph {et~al.}(2016)\citenamefont {Seifert}, \citenamefont {Jaiswal}, \citenamefont {Martens}, \citenamefont {Hannegan}, \citenamefont {Braun}, \citenamefont {Maldonado}, \citenamefont {Freimuth}, \citenamefont {Kronenberg}, \citenamefont {Henrizi}, \citenamefont {Radu}, \citenamefont {Beaurepaire}, \citenamefont {Mokrousov}, \citenamefont {Oppeneer}, \citenamefont {Jourdan}, \citenamefont {Jakob}, \citenamefont {Turchinovich}, \citenamefont {Hayden}, \citenamefont {Wolf}, \citenamefont {M\"unzenberg}, \citenamefont {Kl\"aui},\ and\ \citenamefont {Kampfrath}}]{Seifert2016}%
  \BibitemOpen
  \bibfield  {author} {\bibinfo {author} {\bibfnamefont {T.}~\bibnamefont {Seifert}}, \bibinfo {author} {\bibfnamefont {S.}~\bibnamefont {Jaiswal}}, \bibinfo {author} {\bibfnamefont {U.}~\bibnamefont {Martens}}, \bibinfo {author} {\bibfnamefont {J.}~\bibnamefont {Hannegan}}, \bibinfo {author} {\bibfnamefont {L.}~\bibnamefont {Braun}}, \bibinfo {author} {\bibfnamefont {P.}~\bibnamefont {Maldonado}}, \bibinfo {author} {\bibfnamefont {F.}~\bibnamefont {Freimuth}}, \bibinfo {author} {\bibfnamefont {A.}~\bibnamefont {Kronenberg}}, \bibinfo {author} {\bibfnamefont {J.}~\bibnamefont {Henrizi}}, \bibinfo {author} {\bibfnamefont {I.}~\bibnamefont {Radu}}, \bibinfo {author} {\bibfnamefont {E.}~\bibnamefont {Beaurepaire}}, \bibinfo {author} {\bibfnamefont {Y.}~\bibnamefont {Mokrousov}}, \bibinfo {author} {\bibfnamefont {P.~M.}\ \bibnamefont {Oppeneer}}, \bibinfo {author} {\bibfnamefont {M.}~\bibnamefont {Jourdan}}, \bibinfo {author} {\bibfnamefont {G.}~\bibnamefont {Jakob}}, \bibinfo {author} {\bibfnamefont
  {D.}~\bibnamefont {Turchinovich}}, \bibinfo {author} {\bibfnamefont {L.~M.}\ \bibnamefont {Hayden}}, \bibinfo {author} {\bibfnamefont {M.}~\bibnamefont {Wolf}}, \bibinfo {author} {\bibfnamefont {M.}~\bibnamefont {M\"unzenberg}}, \bibinfo {author} {\bibfnamefont {M.}~\bibnamefont {Kl\"aui}},\ and\ \bibinfo {author} {\bibfnamefont {T.}~\bibnamefont {Kampfrath}},\ }\bibfield  {title} {\bibinfo {title} {Efficient metallic spintronic emitters of ultrabroadband terahertz radiation},\ }\href {https://doi.org/10.1038/nphoton.2016.91} {\bibfield  {journal} {\bibinfo  {journal} {Nature Photonics}\ }\textbf {\bibinfo {volume} {10}},\ \bibinfo {pages} {483} (\bibinfo {year} {2016})}\BibitemShut {NoStop}%
\bibitem [{\citenamefont {Koopmans}\ \emph {et~al.}(2010)\citenamefont {Koopmans}, \citenamefont {Malinowski}, \citenamefont {Longa}, \citenamefont {Steiauf}, \citenamefont {F\"ahnle}, \citenamefont {Roth}, \citenamefont {Cinchetti},\ and\ \citenamefont {Aeschlimann}}]{Koopmans2010}%
  \BibitemOpen
  \bibfield  {author} {\bibinfo {author} {\bibfnamefont {B.}~\bibnamefont {Koopmans}}, \bibinfo {author} {\bibfnamefont {G.}~\bibnamefont {Malinowski}}, \bibinfo {author} {\bibfnamefont {F.~D.}\ \bibnamefont {Longa}}, \bibinfo {author} {\bibfnamefont {D.}~\bibnamefont {Steiauf}}, \bibinfo {author} {\bibfnamefont {M.}~\bibnamefont {F\"ahnle}}, \bibinfo {author} {\bibfnamefont {T.}~\bibnamefont {Roth}}, \bibinfo {author} {\bibfnamefont {M.}~\bibnamefont {Cinchetti}},\ and\ \bibinfo {author} {\bibfnamefont {M.}~\bibnamefont {Aeschlimann}},\ }\bibfield  {title} {\bibinfo {title} {Explaining the paradoxical diversity of ultrafast laser-induced demagnetization},\ }\href {https://doi.org/10.1038/nmat2593} {\bibfield  {journal} {\bibinfo  {journal} {Nature Materials}\ }\textbf {\bibinfo {volume} {9}},\ \bibinfo {pages} {259} (\bibinfo {year} {2010})}\BibitemShut {NoStop}%
\bibitem [{\citenamefont {Beaurepaire}\ \emph {et~al.}(1996)\citenamefont {Beaurepaire}, \citenamefont {Merle}, \citenamefont {Daunois},\ and\ \citenamefont {Bigot}}]{Beaurepaire1996}%
  \BibitemOpen
  \bibfield  {author} {\bibinfo {author} {\bibfnamefont {E.}~\bibnamefont {Beaurepaire}}, \bibinfo {author} {\bibfnamefont {J.-C.}\ \bibnamefont {Merle}}, \bibinfo {author} {\bibfnamefont {A.}~\bibnamefont {Daunois}},\ and\ \bibinfo {author} {\bibfnamefont {J.-Y.}\ \bibnamefont {Bigot}},\ }\bibfield  {title} {\bibinfo {title} {Ultrafast spin dynamics in ferromagnetic nickel},\ }\href {https://doi.org/10.1103/PhysRevLett.76.4250} {\bibfield  {journal} {\bibinfo  {journal} {Physical Review Letters}\ }\textbf {\bibinfo {volume} {76}},\ \bibinfo {pages} {4250} (\bibinfo {year} {1996})}\BibitemShut {NoStop}%
\bibitem [{\citenamefont {Stamm}\ \emph {et~al.}(2007)\citenamefont {Stamm}, \citenamefont {Kachel}, \citenamefont {Pontius}, \citenamefont {Mitzner}, \citenamefont {Quast}, \citenamefont {Holldack}, \citenamefont {Khan}, \citenamefont {Lupulescu}, \citenamefont {Aziz}, \citenamefont {Wietstruk}, \citenamefont {D\"urr},\ and\ \citenamefont {Eberhardt}}]{Stamm2007}%
  \BibitemOpen
  \bibfield  {author} {\bibinfo {author} {\bibfnamefont {C.}~\bibnamefont {Stamm}}, \bibinfo {author} {\bibfnamefont {T.}~\bibnamefont {Kachel}}, \bibinfo {author} {\bibfnamefont {N.}~\bibnamefont {Pontius}}, \bibinfo {author} {\bibfnamefont {R.}~\bibnamefont {Mitzner}}, \bibinfo {author} {\bibfnamefont {T.}~\bibnamefont {Quast}}, \bibinfo {author} {\bibfnamefont {K.}~\bibnamefont {Holldack}}, \bibinfo {author} {\bibfnamefont {S.}~\bibnamefont {Khan}}, \bibinfo {author} {\bibfnamefont {C.}~\bibnamefont {Lupulescu}}, \bibinfo {author} {\bibfnamefont {E.~F.}\ \bibnamefont {Aziz}}, \bibinfo {author} {\bibfnamefont {M.}~\bibnamefont {Wietstruk}}, \bibinfo {author} {\bibfnamefont {H.~A.}\ \bibnamefont {D\"urr}},\ and\ \bibinfo {author} {\bibfnamefont {W.}~\bibnamefont {Eberhardt}},\ }\bibfield  {title} {\bibinfo {title} {Femtosecond modification of electron localization and transfer of angular momentum in nickel},\ }\href {https://doi.org/10.1038/nmat1985} {\bibfield  {journal} {\bibinfo  {journal} {Nature
  Materials}\ }\textbf {\bibinfo {volume} {6}},\ \bibinfo {pages} {740} (\bibinfo {year} {2007})}\BibitemShut {NoStop}%
\bibitem [{\citenamefont {Carpene}\ \emph {et~al.}(2008)\citenamefont {Carpene}, \citenamefont {Mancini}, \citenamefont {Dallera}, \citenamefont {Brenna}, \citenamefont {Puppin},\ and\ \citenamefont {Silvestri}}]{Carpene2008}%
  \BibitemOpen
  \bibfield  {author} {\bibinfo {author} {\bibfnamefont {E.}~\bibnamefont {Carpene}}, \bibinfo {author} {\bibfnamefont {E.}~\bibnamefont {Mancini}}, \bibinfo {author} {\bibfnamefont {C.}~\bibnamefont {Dallera}}, \bibinfo {author} {\bibfnamefont {M.}~\bibnamefont {Brenna}}, \bibinfo {author} {\bibfnamefont {E.}~\bibnamefont {Puppin}},\ and\ \bibinfo {author} {\bibfnamefont {S.~D.}\ \bibnamefont {Silvestri}},\ }\bibfield  {title} {\bibinfo {title} {Dynamics of electron-magnon interaction and ultrafast demagnetization in thin iron films},\ }\href {https://doi.org/10.1103/PhysRevB.78.174422} {\bibfield  {journal} {\bibinfo  {journal} {Physical Review B}\ }\textbf {\bibinfo {volume} {78}},\ \bibinfo {pages} {174422} (\bibinfo {year} {2008})}\BibitemShut {NoStop}%
\bibitem [{\citenamefont {Kimel}\ \emph {et~al.}(2002)\citenamefont {Kimel}, \citenamefont {Pisarev}, \citenamefont {Hohlfeld},\ and\ \citenamefont {Rasing}}]{Kimel-PRL2002}%
  \BibitemOpen
  \bibfield  {author} {\bibinfo {author} {\bibfnamefont {A.~V.}\ \bibnamefont {Kimel}}, \bibinfo {author} {\bibfnamefont {R.~V.}\ \bibnamefont {Pisarev}}, \bibinfo {author} {\bibfnamefont {J.}~\bibnamefont {Hohlfeld}},\ and\ \bibinfo {author} {\bibfnamefont {T.}~\bibnamefont {Rasing}},\ }\bibfield  {title} {\bibinfo {title} {Ultrafast quenching of the antiferromagnetic order in $\mathrm{F}\mathrm{e}\mathrm{B}\mathrm{O}_{\mathrm{3}}$: Direct optical probing of the phonon-magnon coupling},\ }\href {https://doi.org/10.1103/PhysRevLett.89.287401} {\bibfield  {journal} {\bibinfo  {journal} {Phys. Rev. Lett.}\ }\textbf {\bibinfo {volume} {89}},\ \bibinfo {pages} {287401} (\bibinfo {year} {2002})}\BibitemShut {NoStop}%
\bibitem [{\citenamefont {Maehrlein}\ \emph {et~al.}(2018)\citenamefont {Maehrlein}, \citenamefont {Radu}, \citenamefont {Maldonado}, \citenamefont {Paarmann}, \citenamefont {Gensch}, \citenamefont {Kalashnikova}, \citenamefont {Pisarev}, \citenamefont {Wolf}, \citenamefont {Oppeneer}, \citenamefont {Barker},\ and\ \citenamefont {Kampfrath}}]{Maehrlein-SciAdv2018}%
  \BibitemOpen
  \bibfield  {author} {\bibinfo {author} {\bibfnamefont {S.~F.}\ \bibnamefont {Maehrlein}}, \bibinfo {author} {\bibfnamefont {I.}~\bibnamefont {Radu}}, \bibinfo {author} {\bibfnamefont {P.}~\bibnamefont {Maldonado}}, \bibinfo {author} {\bibfnamefont {A.}~\bibnamefont {Paarmann}}, \bibinfo {author} {\bibfnamefont {M.}~\bibnamefont {Gensch}}, \bibinfo {author} {\bibfnamefont {A.~M.}\ \bibnamefont {Kalashnikova}}, \bibinfo {author} {\bibfnamefont {R.~V.}\ \bibnamefont {Pisarev}}, \bibinfo {author} {\bibfnamefont {M.}~\bibnamefont {Wolf}}, \bibinfo {author} {\bibfnamefont {P.~M.}\ \bibnamefont {Oppeneer}}, \bibinfo {author} {\bibfnamefont {J.}~\bibnamefont {Barker}},\ and\ \bibinfo {author} {\bibfnamefont {T.}~\bibnamefont {Kampfrath}},\ }\bibfield  {title} {\bibinfo {title} {Dissecting spin-phonon equilibration in ferrimagnetic insulators by ultrafast lattice excitation},\ }\href {https://doi.org/10.1126/sciadv.aar5164} {\bibfield  {journal} {\bibinfo  {journal} {Science Advances}\ }\textbf {\bibinfo {volume}
  {4}},\ \bibinfo {pages} {eaar5164} (\bibinfo {year} {2018})}\BibitemShut {NoStop}%
\bibitem [{\citenamefont {L{\'o}pez-Flores}\ \emph {et~al.}(2013)\citenamefont {L{\'o}pez-Flores}, \citenamefont {Bergeard}, \citenamefont {Halt{\'e}}, \citenamefont {Stamm}, \citenamefont {Pontius}, \citenamefont {Hehn}, \citenamefont {Otero}, \citenamefont {Beaurepaire},\ and\ \citenamefont {Boeglin}}]{Lopez-PRB2013}%
  \BibitemOpen
  \bibfield  {author} {\bibinfo {author} {\bibfnamefont {V.}~\bibnamefont {L{\'o}pez-Flores}}, \bibinfo {author} {\bibfnamefont {N.}~\bibnamefont {Bergeard}}, \bibinfo {author} {\bibfnamefont {V.}~\bibnamefont {Halt{\'e}}}, \bibinfo {author} {\bibfnamefont {C.}~\bibnamefont {Stamm}}, \bibinfo {author} {\bibfnamefont {N.}~\bibnamefont {Pontius}}, \bibinfo {author} {\bibfnamefont {M.}~\bibnamefont {Hehn}}, \bibinfo {author} {\bibfnamefont {E.}~\bibnamefont {Otero}}, \bibinfo {author} {\bibfnamefont {E.}~\bibnamefont {Beaurepaire}},\ and\ \bibinfo {author} {\bibfnamefont {C.}~\bibnamefont {Boeglin}},\ }\bibfield  {title} {\bibinfo {title} {Role of critical spin fluctuations in ultrafast demagnetization of transition-metal rare-earth alloys},\ }\href {https://doi.org/10.1103/PhysRevB.87.214412} {\bibfield  {journal} {\bibinfo  {journal} {Physical Review B}\ }\textbf {\bibinfo {volume} {87}},\ \bibinfo {pages} {214412} (\bibinfo {year} {2013})}\BibitemShut {NoStop}%
\bibitem [{\citenamefont {Kimling}\ \emph {et~al.}(2014)\citenamefont {Kimling}, \citenamefont {Kimling}, \citenamefont {Wilson}, \citenamefont {Hebler}, \citenamefont {Albrecht},\ and\ \citenamefont {Cahill}}]{Kimling-PRB2014}%
  \BibitemOpen
  \bibfield  {author} {\bibinfo {author} {\bibfnamefont {J.}~\bibnamefont {Kimling}}, \bibinfo {author} {\bibfnamefont {J.}~\bibnamefont {Kimling}}, \bibinfo {author} {\bibfnamefont {R.~B.}\ \bibnamefont {Wilson}}, \bibinfo {author} {\bibfnamefont {B.}~\bibnamefont {Hebler}}, \bibinfo {author} {\bibfnamefont {M.}~\bibnamefont {Albrecht}},\ and\ \bibinfo {author} {\bibfnamefont {D.~G.}\ \bibnamefont {Cahill}},\ }\bibfield  {title} {\bibinfo {title} {Ultrafast demagnetization of {F}e{P}t:{C}u thin films and the role of magnetic heat capacity},\ }\href {https://doi.org/10.1103/PhysRevB.90.224408} {\bibfield  {journal} {\bibinfo  {journal} {Phys. Rev. B}\ }\textbf {\bibinfo {volume} {90}},\ \bibinfo {pages} {224408} (\bibinfo {year} {2014})}\BibitemShut {NoStop}%
\bibitem [{\citenamefont {Susner}\ \emph {et~al.}(2017)\citenamefont {Susner}, \citenamefont {Chyasnavichyus}, \citenamefont {McGuire}, \citenamefont {Ganesh},\ and\ \citenamefont {Maksymovych}}]{Susner2017}%
  \BibitemOpen
  \bibfield  {author} {\bibinfo {author} {\bibfnamefont {M.~A.}\ \bibnamefont {Susner}}, \bibinfo {author} {\bibfnamefont {M.}~\bibnamefont {Chyasnavichyus}}, \bibinfo {author} {\bibfnamefont {M.~A.}\ \bibnamefont {McGuire}}, \bibinfo {author} {\bibfnamefont {P.}~\bibnamefont {Ganesh}},\ and\ \bibinfo {author} {\bibfnamefont {P.}~\bibnamefont {Maksymovych}},\ }\bibfield  {title} {\bibinfo {title} {Metal thio- and selenophosphates as multifunctional van der {W}aals layered materials},\ }\href {https://doi.org/10.1002/adma.201602852} {\bibfield  {journal} {\bibinfo  {journal} {Advanced Materials}\ }\textbf {\bibinfo {volume} {29}},\ \bibinfo {pages} {1602852} (\bibinfo {year} {2017})}\BibitemShut {NoStop}%
\bibitem [{\citenamefont {Wang}\ \emph {et~al.}(2018)\citenamefont {Wang}, \citenamefont {Shifa}, \citenamefont {Yu}, \citenamefont {He}, \citenamefont {Liu}, \citenamefont {Wang}, \citenamefont {Wang}, \citenamefont {Zhan}, \citenamefont {Lou}, \citenamefont {Xia},\ and\ \citenamefont {He}}]{Wang2018}%
  \BibitemOpen
  \bibfield  {author} {\bibinfo {author} {\bibfnamefont {F.}~\bibnamefont {Wang}}, \bibinfo {author} {\bibfnamefont {T.~A.}\ \bibnamefont {Shifa}}, \bibinfo {author} {\bibfnamefont {P.}~\bibnamefont {Yu}}, \bibinfo {author} {\bibfnamefont {P.}~\bibnamefont {He}}, \bibinfo {author} {\bibfnamefont {Y.}~\bibnamefont {Liu}}, \bibinfo {author} {\bibfnamefont {F.}~\bibnamefont {Wang}}, \bibinfo {author} {\bibfnamefont {Z.}~\bibnamefont {Wang}}, \bibinfo {author} {\bibfnamefont {X.}~\bibnamefont {Zhan}}, \bibinfo {author} {\bibfnamefont {X.}~\bibnamefont {Lou}}, \bibinfo {author} {\bibfnamefont {F.}~\bibnamefont {Xia}},\ and\ \bibinfo {author} {\bibfnamefont {J.}~\bibnamefont {He}},\ }\bibfield  {title} {\bibinfo {title} {New frontiers on van der {W}aals layered metal phosphorous trichalcogenides},\ }\href {https://doi.org/10.1002/adfm.201802151} {\bibfield  {journal} {\bibinfo  {journal} {Advanced Functional Materials}\ }\textbf {\bibinfo {volume} {28}},\ \bibinfo {pages} {1802151} (\bibinfo {year}
  {2018})}\BibitemShut {NoStop}%
\bibitem [{\citenamefont {Bellitto}\ \emph {et~al.}(2015)\citenamefont {Bellitto}, \citenamefont {Bauer},\ and\ \citenamefont {Righini}}]{Bellitto2015}%
  \BibitemOpen
  \bibfield  {author} {\bibinfo {author} {\bibfnamefont {C.}~\bibnamefont {Bellitto}}, \bibinfo {author} {\bibfnamefont {E.~M.}\ \bibnamefont {Bauer}},\ and\ \bibinfo {author} {\bibfnamefont {G.}~\bibnamefont {Righini}},\ }\bibfield  {title} {\bibinfo {title} {Organic–inorganic hybrids: From magnetic perovskite metal({II}) halides to multifunctional metal({II}) phosphonates},\ }\href {https://doi.org/10.1016/j.ccr.2014.10.005} {\bibfield  {journal} {\bibinfo  {journal} {Coordination Chemistry Reviews}\ }\textbf {\bibinfo {volume} {289-290}},\ \bibinfo {pages} {123} (\bibinfo {year} {2015})}\BibitemShut {NoStop}%
\bibitem [{\citenamefont {Ingason}\ \emph {et~al.}(2016)\citenamefont {Ingason}, \citenamefont {Dahlqvist},\ and\ \citenamefont {Rosen}}]{Ingason2016}%
  \BibitemOpen
  \bibfield  {author} {\bibinfo {author} {\bibfnamefont {A.~S.}\ \bibnamefont {Ingason}}, \bibinfo {author} {\bibfnamefont {M.}~\bibnamefont {Dahlqvist}},\ and\ \bibinfo {author} {\bibfnamefont {J.}~\bibnamefont {Rosen}},\ }\bibfield  {title} {\bibinfo {title} {Magnetic {MAX} phases from theory and experiments; a review},\ }\href {https://doi.org/10.1088/0953-8984/28/43/433003} {\bibfield  {journal} {\bibinfo  {journal} {Journal of Physics: Condensed Matter}\ }\textbf {\bibinfo {volume} {28}},\ \bibinfo {pages} {433003} (\bibinfo {year} {2016})}\BibitemShut {NoStop}%
\bibitem [{\citenamefont {Jenkins}\ \emph {et~al.}(2022)\citenamefont {Jenkins}, \citenamefont {R{\'o}zsa}, \citenamefont {Atxitia}, \citenamefont {Evans}, \citenamefont {Novoselov},\ and\ \citenamefont {Santos}}]{jenkins_breaking_2022}%
  \BibitemOpen
  \bibfield  {author} {\bibinfo {author} {\bibfnamefont {S.}~\bibnamefont {Jenkins}}, \bibinfo {author} {\bibfnamefont {L.}~\bibnamefont {R{\'o}zsa}}, \bibinfo {author} {\bibfnamefont {U.}~\bibnamefont {Atxitia}}, \bibinfo {author} {\bibfnamefont {R.~F.~L.}\ \bibnamefont {Evans}}, \bibinfo {author} {\bibfnamefont {K.~S.}\ \bibnamefont {Novoselov}},\ and\ \bibinfo {author} {\bibfnamefont {E.~J.~G.}\ \bibnamefont {Santos}},\ }\bibfield  {title} {\bibinfo {title} {Breaking through the {Mermin}-{Wagner} limit in {2D} van der {Waals} magnets},\ }\href {https://doi.org/10.1038/s41467-022-34389-0} {\bibfield  {journal} {\bibinfo  {journal} {Nature Communications}\ }\textbf {\bibinfo {volume} {13}},\ \bibinfo {pages} {6917} (\bibinfo {year} {2022})}\BibitemShut {NoStop}%
\bibitem [{\citenamefont {Kuo}\ \emph {et~al.}(2016)\citenamefont {Kuo}, \citenamefont {Neumann}, \citenamefont {Balamurugan}, \citenamefont {Park}, \citenamefont {Kang}, \citenamefont {Shiu}, \citenamefont {Kang}, \citenamefont {Hong}, \citenamefont {Han}, \citenamefont {Noh} \emph {et~al.}}]{kuo2016exfoliation}%
  \BibitemOpen
  \bibfield  {author} {\bibinfo {author} {\bibfnamefont {C.-T.}\ \bibnamefont {Kuo}}, \bibinfo {author} {\bibfnamefont {M.}~\bibnamefont {Neumann}}, \bibinfo {author} {\bibfnamefont {K.}~\bibnamefont {Balamurugan}}, \bibinfo {author} {\bibfnamefont {H.~J.}\ \bibnamefont {Park}}, \bibinfo {author} {\bibfnamefont {S.}~\bibnamefont {Kang}}, \bibinfo {author} {\bibfnamefont {H.~W.}\ \bibnamefont {Shiu}}, \bibinfo {author} {\bibfnamefont {J.~H.}\ \bibnamefont {Kang}}, \bibinfo {author} {\bibfnamefont {B.~H.}\ \bibnamefont {Hong}}, \bibinfo {author} {\bibfnamefont {M.}~\bibnamefont {Han}}, \bibinfo {author} {\bibfnamefont {T.~W.}\ \bibnamefont {Noh}}, \emph {et~al.},\ }\bibfield  {title} {\bibinfo {title} {Exfoliation and {R}aman spectroscopic fingerprint of few-layer \nips van der {W}aals crystals},\ }\href {https://doi.org/10.1038/srep20904} {\bibfield  {journal} {\bibinfo  {journal} {Sci. rep.}\ }\textbf {\bibinfo {volume} {6}},\ \bibinfo {pages} {20904} (\bibinfo {year} {2016})}\BibitemShut {NoStop}%
\bibitem [{\citenamefont {Xu}\ \emph {et~al.}(2013)\citenamefont {Xu}, \citenamefont {Liang}, \citenamefont {Shi},\ and\ \citenamefont {Chen}}]{Xu_Liang_Shi_Chen_2013}%
  \BibitemOpen
  \bibfield  {author} {\bibinfo {author} {\bibfnamefont {M.}~\bibnamefont {Xu}}, \bibinfo {author} {\bibfnamefont {T.}~\bibnamefont {Liang}}, \bibinfo {author} {\bibfnamefont {M.}~\bibnamefont {Shi}},\ and\ \bibinfo {author} {\bibfnamefont {H.}~\bibnamefont {Chen}},\ }\bibfield  {title} {\bibinfo {title} {Graphene-like two-dimensional materials},\ }\href {https://doi.org/10.1021/cr300263a} {\bibfield  {journal} {\bibinfo  {journal} {Chemical Reviews}\ }\textbf {\bibinfo {volume} {113}},\ \bibinfo {pages} {3766–3798} (\bibinfo {year} {2013})}\BibitemShut {NoStop}%
\bibitem [{\citenamefont {Mas-Ballesté}\ \emph {et~al.}(2011)\citenamefont {Mas-Ballesté}, \citenamefont {Gómez-Navarro}, \citenamefont {Gómez-Herrero},\ and\ \citenamefont {Zamora}}]{Mas-Balleste2011}%
  \BibitemOpen
  \bibfield  {author} {\bibinfo {author} {\bibfnamefont {R.}~\bibnamefont {Mas-Ballesté}}, \bibinfo {author} {\bibfnamefont {C.}~\bibnamefont {Gómez-Navarro}}, \bibinfo {author} {\bibfnamefont {J.}~\bibnamefont {Gómez-Herrero}},\ and\ \bibinfo {author} {\bibfnamefont {F.}~\bibnamefont {Zamora}},\ }\bibfield  {title} {\bibinfo {title} {{2D} materials: to graphene and beyond},\ }\href {https://doi.org/10.1039/C0NR00323A} {\bibfield  {journal} {\bibinfo  {journal} {Nanoscale}\ }\textbf {\bibinfo {volume} {3}},\ \bibinfo {pages} {20} (\bibinfo {year} {2011})}\BibitemShut {NoStop}%
\bibitem [{\citenamefont {Butler}\ \emph {et~al.}(2013)\citenamefont {Butler}, \citenamefont {Hollen}, \citenamefont {Cao}, \citenamefont {Cui}, \citenamefont {Gupta}, \citenamefont {Gutiérrez}, \citenamefont {Heinz}, \citenamefont {Hong}, \citenamefont {Huang}, \citenamefont {Ismach}, \citenamefont {Johnston-Halperin}, \citenamefont {Kuno}, \citenamefont {Plashnitsa}, \citenamefont {Robinson}, \citenamefont {Ruoff}, \citenamefont {Salahuddin}, \citenamefont {Shan}, \citenamefont {Shi}, \citenamefont {Spencer}, \citenamefont {Terrones}, \citenamefont {Windl},\ and\ \citenamefont {Goldberger}}]{Butler2013}%
  \BibitemOpen
  \bibfield  {author} {\bibinfo {author} {\bibfnamefont {S.~Z.}\ \bibnamefont {Butler}}, \bibinfo {author} {\bibfnamefont {S.~M.}\ \bibnamefont {Hollen}}, \bibinfo {author} {\bibfnamefont {L.}~\bibnamefont {Cao}}, \bibinfo {author} {\bibfnamefont {Y.}~\bibnamefont {Cui}}, \bibinfo {author} {\bibfnamefont {J.~A.}\ \bibnamefont {Gupta}}, \bibinfo {author} {\bibfnamefont {H.~R.}\ \bibnamefont {Gutiérrez}}, \bibinfo {author} {\bibfnamefont {T.~F.}\ \bibnamefont {Heinz}}, \bibinfo {author} {\bibfnamefont {S.~S.}\ \bibnamefont {Hong}}, \bibinfo {author} {\bibfnamefont {J.}~\bibnamefont {Huang}}, \bibinfo {author} {\bibfnamefont {A.~F.}\ \bibnamefont {Ismach}}, \bibinfo {author} {\bibfnamefont {E.}~\bibnamefont {Johnston-Halperin}}, \bibinfo {author} {\bibfnamefont {M.}~\bibnamefont {Kuno}}, \bibinfo {author} {\bibfnamefont {V.~V.}\ \bibnamefont {Plashnitsa}}, \bibinfo {author} {\bibfnamefont {R.~D.}\ \bibnamefont {Robinson}}, \bibinfo {author} {\bibfnamefont {R.~S.}\ \bibnamefont {Ruoff}}, \bibinfo {author}
  {\bibfnamefont {S.}~\bibnamefont {Salahuddin}}, \bibinfo {author} {\bibfnamefont {J.}~\bibnamefont {Shan}}, \bibinfo {author} {\bibfnamefont {L.}~\bibnamefont {Shi}}, \bibinfo {author} {\bibfnamefont {M.~G.}\ \bibnamefont {Spencer}}, \bibinfo {author} {\bibfnamefont {M.}~\bibnamefont {Terrones}}, \bibinfo {author} {\bibfnamefont {W.}~\bibnamefont {Windl}},\ and\ \bibinfo {author} {\bibfnamefont {J.~E.}\ \bibnamefont {Goldberger}},\ }\bibfield  {title} {\bibinfo {title} {Progress, challenges, and opportunities in two-dimensional materials beyond graphene},\ }\href {https://doi.org/10.1021/nn400280c} {\bibfield  {journal} {\bibinfo  {journal} {ACS Nano}\ }\textbf {\bibinfo {volume} {7}},\ \bibinfo {pages} {2898} (\bibinfo {year} {2013})}\BibitemShut {NoStop}%
\bibitem [{\citenamefont {Huang}\ \emph {et~al.}(2017)\citenamefont {Huang}, \citenamefont {Clark}, \citenamefont {Navarro-Moratalla}, \citenamefont {Klein}, \citenamefont {Cheng}, \citenamefont {Seyler}, \citenamefont {Zhong}, \citenamefont {Schmidgall}, \citenamefont {McGuire}, \citenamefont {Cobden}, \citenamefont {Yao}, \citenamefont {Xiao}, \citenamefont {Jarillo-Herrero},\ and\ \citenamefont {Xu}}]{Huang2017}%
  \BibitemOpen
  \bibfield  {author} {\bibinfo {author} {\bibfnamefont {B.}~\bibnamefont {Huang}}, \bibinfo {author} {\bibfnamefont {G.}~\bibnamefont {Clark}}, \bibinfo {author} {\bibfnamefont {E.}~\bibnamefont {Navarro-Moratalla}}, \bibinfo {author} {\bibfnamefont {D.~R.}\ \bibnamefont {Klein}}, \bibinfo {author} {\bibfnamefont {R.}~\bibnamefont {Cheng}}, \bibinfo {author} {\bibfnamefont {K.~L.}\ \bibnamefont {Seyler}}, \bibinfo {author} {\bibfnamefont {D.}~\bibnamefont {Zhong}}, \bibinfo {author} {\bibfnamefont {E.}~\bibnamefont {Schmidgall}}, \bibinfo {author} {\bibfnamefont {M.~A.}\ \bibnamefont {McGuire}}, \bibinfo {author} {\bibfnamefont {D.~H.}\ \bibnamefont {Cobden}}, \bibinfo {author} {\bibfnamefont {W.}~\bibnamefont {Yao}}, \bibinfo {author} {\bibfnamefont {D.}~\bibnamefont {Xiao}}, \bibinfo {author} {\bibfnamefont {P.}~\bibnamefont {Jarillo-Herrero}},\ and\ \bibinfo {author} {\bibfnamefont {X.}~\bibnamefont {Xu}},\ }\bibfield  {title} {\bibinfo {title} {Layer-dependent ferromagnetism in a van der {W}aals crystal
  down to the monolayer limit},\ }\href {https://doi.org/10.1038/nature22391} {\bibfield  {journal} {\bibinfo  {journal} {Nature}\ }\textbf {\bibinfo {volume} {546}},\ \bibinfo {pages} {270–273} (\bibinfo {year} {2017})}\BibitemShut {NoStop}%
\bibitem [{\citenamefont {Kim}\ \emph {et~al.}(2019)\citenamefont {Kim}, \citenamefont {Lim},\ and\ \citenamefont {Lee}}]{Kim2019}%
  \BibitemOpen
  \bibfield  {author} {\bibinfo {author} {\bibfnamefont {K.}~\bibnamefont {Kim}}, \bibinfo {author} {\bibfnamefont {S.~Y.}\ \bibnamefont {Lim}},\ and\ \bibinfo {author} {\bibfnamefont {J.~U.}\ \bibnamefont {Lee}},\ }\bibfield  {title} {\bibinfo {title} {{Suppression of magnetic ordering in {XXZ}-type antiferromagnetic monolayer \nips}},\ }\bibfield  {journal} {\bibinfo  {journal} {Nature Communications}\ }\textbf {\bibinfo {volume} {10}},\ \href {https://doi.org/10.1038/s41467-018-08284-6} {10.1038/s41467-018-08284-6} (\bibinfo {year} {2019})\BibitemShut {NoStop}%
\bibitem [{\citenamefont {Kim}\ and\ \citenamefont {Park}(2021)}]{Kim_Park_2021}%
  \BibitemOpen
  \bibfield  {author} {\bibinfo {author} {\bibfnamefont {T.~Y.}\ \bibnamefont {Kim}}\ and\ \bibinfo {author} {\bibfnamefont {C.-H.}\ \bibnamefont {Park}},\ }\bibfield  {title} {\bibinfo {title} {Magnetic anisotropy and magnetic ordering of transition-metal phosphorus trisulfides},\ }\href {https://doi.org/10.1021/acs.nanolett.1c03992} {\bibfield  {journal} {\bibinfo  {journal} {Nano Letters}\ }\textbf {\bibinfo {volume} {21}},\ \bibinfo {pages} {10114} (\bibinfo {year} {2021})}\BibitemShut {NoStop}%
\bibitem [{\citenamefont {Caretta}\ \emph {et~al.}(2015)\citenamefont {Caretta}, \citenamefont {Donker}, \citenamefont {Polyakov}, \citenamefont {Palstra},\ and\ \citenamefont {van Loosdrecht}}]{Caretta-PRB2015}%
  \BibitemOpen
  \bibfield  {author} {\bibinfo {author} {\bibfnamefont {A.}~\bibnamefont {Caretta}}, \bibinfo {author} {\bibfnamefont {M.~C.}\ \bibnamefont {Donker}}, \bibinfo {author} {\bibfnamefont {A.~O.}\ \bibnamefont {Polyakov}}, \bibinfo {author} {\bibfnamefont {T.~T.~M.}\ \bibnamefont {Palstra}},\ and\ \bibinfo {author} {\bibfnamefont {P.~H.~M.}\ \bibnamefont {van Loosdrecht}},\ }\bibfield  {title} {\bibinfo {title} {Photoinduced magnetization enhancement in two-dimensional weakly anisotropic {H}eisenberg magnets},\ }\href {https://doi.org/10.1103/PhysRevB.91.020405} {\bibfield  {journal} {\bibinfo  {journal} {Phys. Rev. B}\ }\textbf {\bibinfo {volume} {91}},\ \bibinfo {pages} {020405} (\bibinfo {year} {2015})}\BibitemShut {NoStop}%
\bibitem [{\citenamefont {Mertens}\ \emph {et~al.}(2023)\citenamefont {Mertens}, \citenamefont {M\"ankeb\"uscher}, \citenamefont {Parlak}, \citenamefont {Boix-Constant}, \citenamefont {Ma{\~n}as-Valero}, \citenamefont {Matzer}, \citenamefont {Adhikari}, \citenamefont {Bonanni}, \citenamefont {Coronado}, \citenamefont {Kalashnikova}, \citenamefont {Bossini},\ and\ \citenamefont {Cinchetti}}]{Mertens2022}%
  \BibitemOpen
  \bibfield  {author} {\bibinfo {author} {\bibfnamefont {F.}~\bibnamefont {Mertens}}, \bibinfo {author} {\bibfnamefont {D.}~\bibnamefont {M\"ankeb\"uscher}}, \bibinfo {author} {\bibfnamefont {U.}~\bibnamefont {Parlak}}, \bibinfo {author} {\bibfnamefont {C.}~\bibnamefont {Boix-Constant}}, \bibinfo {author} {\bibfnamefont {S.}~\bibnamefont {Ma{\~n}as-Valero}}, \bibinfo {author} {\bibfnamefont {M.}~\bibnamefont {Matzer}}, \bibinfo {author} {\bibfnamefont {R.}~\bibnamefont {Adhikari}}, \bibinfo {author} {\bibfnamefont {A.}~\bibnamefont {Bonanni}}, \bibinfo {author} {\bibfnamefont {E.}~\bibnamefont {Coronado}}, \bibinfo {author} {\bibfnamefont {A.~M.}\ \bibnamefont {Kalashnikova}}, \bibinfo {author} {\bibfnamefont {D.}~\bibnamefont {Bossini}},\ and\ \bibinfo {author} {\bibfnamefont {M.}~\bibnamefont {Cinchetti}},\ }\bibfield  {title} {\bibinfo {title} {Ultrafast coherent {TH}z lattice dynamics coupled to spins in the van der {W}aals antiferromagnet \feps},\ }\href {https://doi.org/10.1002/adma.202208355}
  {\bibfield  {journal} {\bibinfo  {journal} {Advanced Materials}\ }\textbf {\bibinfo {volume} {35}},\ \bibinfo {pages} {2208355} (\bibinfo {year} {2023})}\BibitemShut {NoStop}%
\bibitem [{\citenamefont {Afanasiev}\ \emph {et~al.}(2021)\citenamefont {Afanasiev}, \citenamefont {Hortensius}, \citenamefont {Matthiesen}, \citenamefont {Mañas-Valero}, \citenamefont {Šiškins}, \citenamefont {Lee}, \citenamefont {Lesne}, \citenamefont {van~der Zant}, \citenamefont {Steeneken}, \citenamefont {Ivanov}, \citenamefont {Coronado},\ and\ \citenamefont {Caviglia}}]{Afanasiev2021}%
  \BibitemOpen
  \bibfield  {author} {\bibinfo {author} {\bibfnamefont {D.}~\bibnamefont {Afanasiev}}, \bibinfo {author} {\bibfnamefont {J.~R.}\ \bibnamefont {Hortensius}}, \bibinfo {author} {\bibfnamefont {M.}~\bibnamefont {Matthiesen}}, \bibinfo {author} {\bibfnamefont {S.}~\bibnamefont {Mañas-Valero}}, \bibinfo {author} {\bibfnamefont {M.}~\bibnamefont {Šiškins}}, \bibinfo {author} {\bibfnamefont {M.}~\bibnamefont {Lee}}, \bibinfo {author} {\bibfnamefont {E.}~\bibnamefont {Lesne}}, \bibinfo {author} {\bibfnamefont {H.~S.~J.}\ \bibnamefont {van~der Zant}}, \bibinfo {author} {\bibfnamefont {P.~G.}\ \bibnamefont {Steeneken}}, \bibinfo {author} {\bibfnamefont {B.~A.}\ \bibnamefont {Ivanov}}, \bibinfo {author} {\bibfnamefont {E.}~\bibnamefont {Coronado}},\ and\ \bibinfo {author} {\bibfnamefont {A.~D.}\ \bibnamefont {Caviglia}},\ }\bibfield  {title} {\bibinfo {title} {Controlling the anisotropy of a van der {W}aals antiferromagnet with light},\ }\href {https://doi.org/10.1126/sciadv.abf3096} {\bibfield  {journal} {\bibinfo
  {journal} {Science Advances}\ }\textbf {\bibinfo {volume} {7}},\ \bibinfo {pages} {eabf3096} (\bibinfo {year} {2021})}\BibitemShut {NoStop}%
\bibitem [{\citenamefont {Zhou}\ \emph {et~al.}(2022)\citenamefont {Zhou}, \citenamefont {Hwangbo}, \citenamefont {Zhang}, \citenamefont {Wang}, \citenamefont {Shen}, \citenamefont {Zhang}, \citenamefont {Jiang}, \citenamefont {Zong}, \citenamefont {Su}, \citenamefont {Zajac}, \citenamefont {Ahn}, \citenamefont {Walko}, \citenamefont {Schaller}, \citenamefont {Chu}, \citenamefont {Gedik}, \citenamefont {Xu}, \citenamefont {Xiao},\ and\ \citenamefont {Wen}}]{Zhou2022}%
  \BibitemOpen
  \bibfield  {author} {\bibinfo {author} {\bibfnamefont {F.}~\bibnamefont {Zhou}}, \bibinfo {author} {\bibfnamefont {K.}~\bibnamefont {Hwangbo}}, \bibinfo {author} {\bibfnamefont {Q.}~\bibnamefont {Zhang}}, \bibinfo {author} {\bibfnamefont {C.}~\bibnamefont {Wang}}, \bibinfo {author} {\bibfnamefont {L.}~\bibnamefont {Shen}}, \bibinfo {author} {\bibfnamefont {J.}~\bibnamefont {Zhang}}, \bibinfo {author} {\bibfnamefont {Q.}~\bibnamefont {Jiang}}, \bibinfo {author} {\bibfnamefont {A.}~\bibnamefont {Zong}}, \bibinfo {author} {\bibfnamefont {Y.}~\bibnamefont {Su}}, \bibinfo {author} {\bibfnamefont {M.}~\bibnamefont {Zajac}}, \bibinfo {author} {\bibfnamefont {Y.}~\bibnamefont {Ahn}}, \bibinfo {author} {\bibfnamefont {D.~A.}\ \bibnamefont {Walko}}, \bibinfo {author} {\bibfnamefont {R.~D.}\ \bibnamefont {Schaller}}, \bibinfo {author} {\bibfnamefont {J.-H.}\ \bibnamefont {Chu}}, \bibinfo {author} {\bibfnamefont {N.}~\bibnamefont {Gedik}}, \bibinfo {author} {\bibfnamefont {X.}~\bibnamefont {Xu}}, \bibinfo {author}
  {\bibfnamefont {D.}~\bibnamefont {Xiao}},\ and\ \bibinfo {author} {\bibfnamefont {H.}~\bibnamefont {Wen}},\ }\bibfield  {title} {\bibinfo {title} {Dynamical criticality of spin-shear coupling in van der {W}aals antiferromagnets},\ }\href {https://doi.org/10.1038/s41467-022-34376-5} {\bibfield  {journal} {\bibinfo  {journal} {Nature Communications}\ }\textbf {\bibinfo {volume} {13}},\ \bibinfo {pages} {6598} (\bibinfo {year} {2022})}\BibitemShut {NoStop}%
\bibitem [{\citenamefont {Khusyainov}\ \emph {et~al.}(2023)\citenamefont {Khusyainov}, \citenamefont {Gareev}, \citenamefont {Radovskaia}, \citenamefont {Sampathkumar}, \citenamefont {Acharya}, \citenamefont {Šiškins}, \citenamefont {Mañas-Valero}, \citenamefont {Ivanov}, \citenamefont {Coronado}, \citenamefont {Rasing}, \citenamefont {Kimel},\ and\ \citenamefont {Afanasiev}}]{Khusyainov2023}%
  \BibitemOpen
  \bibfield  {author} {\bibinfo {author} {\bibfnamefont {D.}~\bibnamefont {Khusyainov}}, \bibinfo {author} {\bibfnamefont {T.}~\bibnamefont {Gareev}}, \bibinfo {author} {\bibfnamefont {V.}~\bibnamefont {Radovskaia}}, \bibinfo {author} {\bibfnamefont {K.}~\bibnamefont {Sampathkumar}}, \bibinfo {author} {\bibfnamefont {S.}~\bibnamefont {Acharya}}, \bibinfo {author} {\bibfnamefont {M.}~\bibnamefont {Šiškins}}, \bibinfo {author} {\bibfnamefont {S.}~\bibnamefont {Mañas-Valero}}, \bibinfo {author} {\bibfnamefont {B.~A.}\ \bibnamefont {Ivanov}}, \bibinfo {author} {\bibfnamefont {E.}~\bibnamefont {Coronado}}, \bibinfo {author} {\bibfnamefont {T.}~\bibnamefont {Rasing}}, \bibinfo {author} {\bibfnamefont {A.~V.}\ \bibnamefont {Kimel}},\ and\ \bibinfo {author} {\bibfnamefont {D.}~\bibnamefont {Afanasiev}},\ }\bibfield  {title} {\bibinfo {title} {Ultrafast laser-induced spin–lattice dynamics in the van der {Waals} antiferromagnet {CoPS3}},\ }\href {https://doi.org/10.1063/5.0146128} {\bibfield  {journal} {\bibinfo
  {journal} {APL Materials}\ }\textbf {\bibinfo {volume} {11}},\ \bibinfo {pages} {071104} (\bibinfo {year} {2023})}\BibitemShut {NoStop}%
\bibitem [{\citenamefont {Lichtenberg}\ \emph {et~al.}(2022)\citenamefont {Lichtenberg}, \citenamefont {Schippers}, \citenamefont {van Kooten}, \citenamefont {Evers}, \citenamefont {Barcones}, \citenamefont {Guimarães},\ and\ \citenamefont {Koopmans}}]{Lichtenberg-2D2023}%
  \BibitemOpen
  \bibfield  {author} {\bibinfo {author} {\bibfnamefont {T.}~\bibnamefont {Lichtenberg}}, \bibinfo {author} {\bibfnamefont {C.~F.}\ \bibnamefont {Schippers}}, \bibinfo {author} {\bibfnamefont {S.~C.~P.}\ \bibnamefont {van Kooten}}, \bibinfo {author} {\bibfnamefont {S.~G.~F.}\ \bibnamefont {Evers}}, \bibinfo {author} {\bibfnamefont {B.}~\bibnamefont {Barcones}}, \bibinfo {author} {\bibfnamefont {M.~H.~D.}\ \bibnamefont {Guimarães}},\ and\ \bibinfo {author} {\bibfnamefont {B.}~\bibnamefont {Koopmans}},\ }\bibfield  {title} {\bibinfo {title} {Anisotropic laser-pulse-induced magnetization dynamics in van der {W}aals magnet {F}e$_3${G}e{T}e$_2$},\ }\href {https://doi.org/10.1088/2053-1583/ac9dab} {\bibfield  {journal} {\bibinfo  {journal} {2D Materials}\ }\textbf {\bibinfo {volume} {10}},\ \bibinfo {pages} {015008} (\bibinfo {year} {2022})}\BibitemShut {NoStop}%
\bibitem [{\citenamefont {Zhang}\ \emph {et~al.}(2021{\natexlab{a}})\citenamefont {Zhang}, \citenamefont {Jiang}, \citenamefont {Lee}, \citenamefont {Lee}, \citenamefont {Mak},\ and\ \citenamefont {Shan}}]{Zhang2021}%
  \BibitemOpen
  \bibfield  {author} {\bibinfo {author} {\bibfnamefont {X.-X.}\ \bibnamefont {Zhang}}, \bibinfo {author} {\bibfnamefont {S.}~\bibnamefont {Jiang}}, \bibinfo {author} {\bibfnamefont {J.}~\bibnamefont {Lee}}, \bibinfo {author} {\bibfnamefont {C.}~\bibnamefont {Lee}}, \bibinfo {author} {\bibfnamefont {K.~F.}\ \bibnamefont {Mak}},\ and\ \bibinfo {author} {\bibfnamefont {J.}~\bibnamefont {Shan}},\ }\bibfield  {title} {\bibinfo {title} {Spin dynamics slowdown near the antiferromagnetic critical point in atomically thin \feps},\ }\href {https://doi.org/10.1021/acs.nanolett.1c00870} {\bibfield  {journal} {\bibinfo  {journal} {Nano Letters}\ }\textbf {\bibinfo {volume} {21}},\ \bibinfo {pages} {5045} (\bibinfo {year} {2021}{\natexlab{a}})}\BibitemShut {NoStop}%
\bibitem [{\citenamefont {Ouvrard}\ \emph {et~al.}(1985)\citenamefont {Ouvrard}, \citenamefont {Brec},\ and\ \citenamefont {Rouxel}}]{Ouvrard1985}%
  \BibitemOpen
  \bibfield  {author} {\bibinfo {author} {\bibfnamefont {G.}~\bibnamefont {Ouvrard}}, \bibinfo {author} {\bibfnamefont {R.}~\bibnamefont {Brec}},\ and\ \bibinfo {author} {\bibfnamefont {J.}~\bibnamefont {Rouxel}},\ }\bibfield  {title} {\bibinfo {title} {Structural determination of some \mps layered phases ({M = Mn, Fe, Co, Ni and Cd})},\ }\href {https://doi.org/10.1016/0025-5408(85)90092-3} {\bibfield  {journal} {\bibinfo  {journal} {Materials Research Bulletin}\ }\textbf {\bibinfo {volume} {20}},\ \bibinfo {pages} {1181} (\bibinfo {year} {1985})}\BibitemShut {NoStop}%
\bibitem [{\citenamefont {Joy}\ and\ \citenamefont {Vasudevan}(1992{\natexlab{a}})}]{JoyVasudevan1992}%
  \BibitemOpen
  \bibfield  {author} {\bibinfo {author} {\bibfnamefont {P.~A.}\ \bibnamefont {Joy}}\ and\ \bibinfo {author} {\bibfnamefont {S.}~\bibnamefont {Vasudevan}},\ }\bibfield  {title} {\bibinfo {title} {Magnetism in the layered transition-metal thiophosphates \textsc{MPS}$_{3}$ (\textsc{M}=\textsc{M}n, \textsc{F}e, \textsc{N}i)},\ }\href {https://doi.org/10.1103/PhysRevB.46.5425} {\bibfield  {journal} {\bibinfo  {journal} {Phys. Rev. B}\ }\textbf {\bibinfo {volume} {46}},\ \bibinfo {pages} {5425} (\bibinfo {year} {1992}{\natexlab{a}})}\BibitemShut {NoStop}%
\bibitem [{\citenamefont {Wildes}\ \emph {et~al.}(2015)\citenamefont {Wildes}, \citenamefont {Simonet}, \citenamefont {Ressouche}, \citenamefont {McIntyre}, \citenamefont {Avdeev}, \citenamefont {Suard}, \citenamefont {Kimber}, \citenamefont {Lan\ifmmode~\mbox{\c{c}}\else \c{c}\fi{}on}, \citenamefont {Pepe}, \citenamefont {Moubaraki},\ and\ \citenamefont {Hicks}}]{Wildes_PRB2015}%
  \BibitemOpen
  \bibfield  {author} {\bibinfo {author} {\bibfnamefont {A.~R.}\ \bibnamefont {Wildes}}, \bibinfo {author} {\bibfnamefont {V.}~\bibnamefont {Simonet}}, \bibinfo {author} {\bibfnamefont {E.}~\bibnamefont {Ressouche}}, \bibinfo {author} {\bibfnamefont {G.~J.}\ \bibnamefont {McIntyre}}, \bibinfo {author} {\bibfnamefont {M.}~\bibnamefont {Avdeev}}, \bibinfo {author} {\bibfnamefont {E.}~\bibnamefont {Suard}}, \bibinfo {author} {\bibfnamefont {S.~A.~J.}\ \bibnamefont {Kimber}}, \bibinfo {author} {\bibfnamefont {D.}~\bibnamefont {Lan\ifmmode~\mbox{\c{c}}\else \c{c}\fi{}on}}, \bibinfo {author} {\bibfnamefont {G.}~\bibnamefont {Pepe}}, \bibinfo {author} {\bibfnamefont {B.}~\bibnamefont {Moubaraki}},\ and\ \bibinfo {author} {\bibfnamefont {T.~J.}\ \bibnamefont {Hicks}},\ }\bibfield  {title} {\bibinfo {title} {{Magnetic structure of the quasi-two-dimensional antiferromagnet \nips}},\ }\href {https://doi.org/10.1103/PhysRevB.92.224408} {\bibfield  {journal} {\bibinfo  {journal} {Phys. Rev. B}\ }\textbf {\bibinfo {volume}
  {92}},\ \bibinfo {pages} {224408} (\bibinfo {year} {2015})}\BibitemShut {NoStop}%
\bibitem [{\citenamefont {Brec}\ \emph {et~al.}(1979)\citenamefont {Brec}, \citenamefont {Schleich}, \citenamefont {Ouvrard}, \citenamefont {Louisy},\ and\ \citenamefont {Rouxel}}]{Brec1979}%
  \BibitemOpen
  \bibfield  {author} {\bibinfo {author} {\bibfnamefont {R.}~\bibnamefont {Brec}}, \bibinfo {author} {\bibfnamefont {D.~M.}\ \bibnamefont {Schleich}}, \bibinfo {author} {\bibfnamefont {G.}~\bibnamefont {Ouvrard}}, \bibinfo {author} {\bibfnamefont {A.}~\bibnamefont {Louisy}},\ and\ \bibinfo {author} {\bibfnamefont {J.}~\bibnamefont {Rouxel}},\ }\bibfield  {title} {\bibinfo {title} {Physical properties of lithium intercalation compounds of the layered transition chalcogenophosphates},\ }\bibfield  {journal} {\bibinfo  {journal} {Inorganic Chemistry}\ }\textbf {\bibinfo {volume} {18}},\ \href {https://doi.org/10.1021/ic50197a018} {10.1021/ic50197a018} (\bibinfo {year} {1979})\BibitemShut {NoStop}%
\bibitem [{\citenamefont {Grasso}\ \emph {et~al.}(1986)\citenamefont {Grasso}, \citenamefont {Santangelo},\ and\ \citenamefont {Piacentini}}]{Grasso-SSI1986}%
  \BibitemOpen
  \bibfield  {author} {\bibinfo {author} {\bibfnamefont {V.}~\bibnamefont {Grasso}}, \bibinfo {author} {\bibfnamefont {S.}~\bibnamefont {Santangelo}},\ and\ \bibinfo {author} {\bibfnamefont {M.}~\bibnamefont {Piacentini}},\ }\bibfield  {title} {\bibinfo {title} {Optical absorption spectra of some transition metal thiophosphates},\ }\href {https://doi.org/10.1016/0167-2738(86)90028-7} {\bibfield  {journal} {\bibinfo  {journal} {Solid State Ionics}\ }\textbf {\bibinfo {volume} {20}},\ \bibinfo {pages} {9} (\bibinfo {year} {1986})}\BibitemShut {NoStop}%
\bibitem [{\citenamefont {Kim}\ \emph {et~al.}(2023)\citenamefont {Kim}, \citenamefont {Huang}, \citenamefont {Guo}, \citenamefont {Li}, \citenamefont {Rocca}, \citenamefont {Gao}, \citenamefont {Choe}, \citenamefont {Lujan}, \citenamefont {Wu}, \citenamefont {Lin}, \citenamefont {Baldini}, \citenamefont {Yang}, \citenamefont {Sharma}, \citenamefont {Kalaivanan}, \citenamefont {Sankar}, \citenamefont {Lee}, \citenamefont {Ping},\ and\ \citenamefont {Li}}]{Kim-AdvMater2023}%
  \BibitemOpen
  \bibfield  {author} {\bibinfo {author} {\bibfnamefont {D.~S.}\ \bibnamefont {Kim}}, \bibinfo {author} {\bibfnamefont {D.}~\bibnamefont {Huang}}, \bibinfo {author} {\bibfnamefont {C.}~\bibnamefont {Guo}}, \bibinfo {author} {\bibfnamefont {K.}~\bibnamefont {Li}}, \bibinfo {author} {\bibfnamefont {D.}~\bibnamefont {Rocca}}, \bibinfo {author} {\bibfnamefont {F.~Y.}\ \bibnamefont {Gao}}, \bibinfo {author} {\bibfnamefont {J.}~\bibnamefont {Choe}}, \bibinfo {author} {\bibfnamefont {D.}~\bibnamefont {Lujan}}, \bibinfo {author} {\bibfnamefont {T.}~\bibnamefont {Wu}}, \bibinfo {author} {\bibfnamefont {K.}~\bibnamefont {Lin}}, \bibinfo {author} {\bibfnamefont {E.}~\bibnamefont {Baldini}}, \bibinfo {author} {\bibfnamefont {L.}~\bibnamefont {Yang}}, \bibinfo {author} {\bibfnamefont {S.}~\bibnamefont {Sharma}}, \bibinfo {author} {\bibfnamefont {R.}~\bibnamefont {Kalaivanan}}, \bibinfo {author} {\bibfnamefont {R.}~\bibnamefont {Sankar}}, \bibinfo {author} {\bibfnamefont {S.}~\bibnamefont {Lee}}, \bibinfo {author}
  {\bibfnamefont {Y.}~\bibnamefont {Ping}},\ and\ \bibinfo {author} {\bibfnamefont {X.}~\bibnamefont {Li}},\ }\bibfield  {title} {\bibinfo {title} {Anisotropic excitons reveal local spin chain directions in a van der {W}aals antiferromagnet},\ }\href {https://doi.org/10.1002/adma.202206585} {\bibfield  {journal} {\bibinfo  {journal} {Advanced Materials}\ ,\ \bibinfo {pages} {2206585}} (\bibinfo {year} {2023})}\BibitemShut {NoStop}%
\bibitem [{\citenamefont {Lan{\c{c}}on}\ \emph {et~al.}(2016)\citenamefont {Lan{\c{c}}on}, \citenamefont {Walker}, \citenamefont {Ressouche}, \citenamefont {Ouladdiaf}, \citenamefont {Rule}, \citenamefont {McIntyre}, \citenamefont {Hicks}, \citenamefont {R{\o}nnow},\ and\ \citenamefont {Wildes}}]{Lancon2016}%
  \BibitemOpen
  \bibfield  {author} {\bibinfo {author} {\bibfnamefont {D.}~\bibnamefont {Lan{\c{c}}on}}, \bibinfo {author} {\bibfnamefont {H.~C.}\ \bibnamefont {Walker}}, \bibinfo {author} {\bibfnamefont {E.}~\bibnamefont {Ressouche}}, \bibinfo {author} {\bibfnamefont {B.}~\bibnamefont {Ouladdiaf}}, \bibinfo {author} {\bibfnamefont {K.~C.}\ \bibnamefont {Rule}}, \bibinfo {author} {\bibfnamefont {G.~J.}\ \bibnamefont {McIntyre}}, \bibinfo {author} {\bibfnamefont {T.~J.}\ \bibnamefont {Hicks}}, \bibinfo {author} {\bibfnamefont {H.~M.}\ \bibnamefont {R{\o}nnow}},\ and\ \bibinfo {author} {\bibfnamefont {A.~R.}\ \bibnamefont {Wildes}},\ }\bibfield  {title} {\bibinfo {title} {{Magnetic structure and magnon dynamics of the quasi-two-dimensional antiferromagnet \feps}},\ }\href {https://doi.org/10.1103/PhysRevB.94.214407} {\bibfield  {journal} {\bibinfo  {journal} {Physical Review B}\ }\textbf {\bibinfo {volume} {94}},\ \bibinfo {pages} {1} (\bibinfo {year} {2016})}\BibitemShut {NoStop}%
\bibitem [{\citenamefont {Lee}\ \emph {et~al.}(2016)\citenamefont {Lee}, \citenamefont {Lee}, \citenamefont {Ryoo}, \citenamefont {Kang}, \citenamefont {Kim}, \citenamefont {Kim}, \citenamefont {Park}, \citenamefont {Park},\ and\ \citenamefont {Cheong}}]{Lee2016}%
  \BibitemOpen
  \bibfield  {author} {\bibinfo {author} {\bibfnamefont {J.-U.}\ \bibnamefont {Lee}}, \bibinfo {author} {\bibfnamefont {S.}~\bibnamefont {Lee}}, \bibinfo {author} {\bibfnamefont {J.~H.}\ \bibnamefont {Ryoo}}, \bibinfo {author} {\bibfnamefont {S.}~\bibnamefont {Kang}}, \bibinfo {author} {\bibfnamefont {T.~Y.}\ \bibnamefont {Kim}}, \bibinfo {author} {\bibfnamefont {P.}~\bibnamefont {Kim}}, \bibinfo {author} {\bibfnamefont {C.-H.}\ \bibnamefont {Park}}, \bibinfo {author} {\bibfnamefont {J.-G.}\ \bibnamefont {Park}},\ and\ \bibinfo {author} {\bibfnamefont {H.}~\bibnamefont {Cheong}},\ }\bibfield  {title} {\bibinfo {title} {Ising-type magnetic ordering in atomically thin \feps},\ }\href {https://doi.org/10.1021/acs.nanolett.6b03052} {\bibfield  {journal} {\bibinfo  {journal} {Nano Letters}\ }\textbf {\bibinfo {volume} {16}},\ \bibinfo {pages} {7433} (\bibinfo {year} {2016})}\BibitemShut {NoStop}%
\bibitem [{\citenamefont {Momma}\ and\ \citenamefont {Izumi}(2011)}]{momma_vesta3_threedimensional_2011}%
  \BibitemOpen
  \bibfield  {author} {\bibinfo {author} {\bibfnamefont {K.}~\bibnamefont {Momma}}\ and\ \bibinfo {author} {\bibfnamefont {F.}~\bibnamefont {Izumi}},\ }\bibfield  {title} {\bibinfo {title} {{{VESTA3}} for three-dimensional visualization of crystal, volumetric and morphology data},\ }\href {https://doi.org/10.1107/S0021889811038970} {\bibfield  {journal} {\bibinfo  {journal} {Journal of Applied Crystallography}\ }\textbf {\bibinfo {volume} {44}},\ \bibinfo {pages} {1272} (\bibinfo {year} {2011})}\BibitemShut {NoStop}%
\bibitem [{\citenamefont {\v{S}i\v{s}kins}\ \emph {et~al.}(2020)\citenamefont {\v{S}i\v{s}kins}, \citenamefont {Lee}, \citenamefont {Ma{\~n}as-Valero}, \citenamefont {Coronado}, \citenamefont {Blanter}, \citenamefont {van~der Zant},\ and\ \citenamefont {Steeneken}}]{Siskins2020}%
  \BibitemOpen
  \bibfield  {author} {\bibinfo {author} {\bibfnamefont {M.}~\bibnamefont {\v{S}i\v{s}kins}}, \bibinfo {author} {\bibfnamefont {M.}~\bibnamefont {Lee}}, \bibinfo {author} {\bibfnamefont {S.}~\bibnamefont {Ma{\~n}as-Valero}}, \bibinfo {author} {\bibfnamefont {E.}~\bibnamefont {Coronado}}, \bibinfo {author} {\bibfnamefont {Y.~M.}\ \bibnamefont {Blanter}}, \bibinfo {author} {\bibfnamefont {H.~S.}\ \bibnamefont {van~der Zant}},\ and\ \bibinfo {author} {\bibfnamefont {P.~G.}\ \bibnamefont {Steeneken}},\ }\bibfield  {title} {\bibinfo {title} {Magnetic and electronic phase transitions probed by nanomechanical resonators},\ }\bibfield  {journal} {\bibinfo  {journal} {Nature Communications}\ }\textbf {\bibinfo {volume} {11}},\ \href {https://doi.org/10.1038/s41467-020-16430-2} {10.1038/s41467-020-16430-2} (\bibinfo {year} {2020})\BibitemShut {NoStop}%
\bibitem [{\citenamefont {Liu}\ \emph {et~al.}(2021)\citenamefont {Liu}, \citenamefont {Granados~del \'Aguila}, \citenamefont {Bhowmick}, \citenamefont {Gan}, \citenamefont {Thu Ha~Do}, \citenamefont {Prosnikov}, \citenamefont {Sedmidubsk\'y}, \citenamefont {Sofer}, \citenamefont {Christianen}, \citenamefont {Sengupta},\ and\ \citenamefont {Xiong}}]{Liu-PRL2021}%
  \BibitemOpen
  \bibfield  {author} {\bibinfo {author} {\bibfnamefont {S.}~\bibnamefont {Liu}}, \bibinfo {author} {\bibfnamefont {A.}~\bibnamefont {Granados~del \'Aguila}}, \bibinfo {author} {\bibfnamefont {D.}~\bibnamefont {Bhowmick}}, \bibinfo {author} {\bibfnamefont {C.~K.}\ \bibnamefont {Gan}}, \bibinfo {author} {\bibfnamefont {T.}~\bibnamefont {Thu Ha~Do}}, \bibinfo {author} {\bibfnamefont {M.~A.}\ \bibnamefont {Prosnikov}}, \bibinfo {author} {\bibfnamefont {D.}~\bibnamefont {Sedmidubsk\'y}}, \bibinfo {author} {\bibfnamefont {Z.}~\bibnamefont {Sofer}}, \bibinfo {author} {\bibfnamefont {P.~C.~M.}\ \bibnamefont {Christianen}}, \bibinfo {author} {\bibfnamefont {P.}~\bibnamefont {Sengupta}},\ and\ \bibinfo {author} {\bibfnamefont {Q.}~\bibnamefont {Xiong}},\ }\bibfield  {title} {\bibinfo {title} {Direct observation of magnon-phonon strong coupling in two-dimensional antiferromagnet at high magnetic fields},\ }\href {https://doi.org/10.1103/PhysRevLett.127.097401} {\bibfield  {journal} {\bibinfo  {journal} {Phys. Rev.
  Lett.}\ }\textbf {\bibinfo {volume} {127}},\ \bibinfo {pages} {097401} (\bibinfo {year} {2021})}\BibitemShut {NoStop}%
\bibitem [{\citenamefont {Vaz}\ \emph {et~al.}(2008)\citenamefont {Vaz}, \citenamefont {Bland},\ and\ \citenamefont {Lauhoff}}]{Vaz2008}%
  \BibitemOpen
  \bibfield  {author} {\bibinfo {author} {\bibfnamefont {C.~A.}\ \bibnamefont {Vaz}}, \bibinfo {author} {\bibfnamefont {J.~A.}\ \bibnamefont {Bland}},\ and\ \bibinfo {author} {\bibfnamefont {G.}~\bibnamefont {Lauhoff}},\ }\bibfield  {title} {\bibinfo {title} {Magnetism in ultrathin film structures},\ }\href {https://doi.org/10.1088/0034-4885/71/5/056501} {\bibfield  {journal} {\bibinfo  {journal} {Reports on Progress in Physics}\ }\textbf {\bibinfo {volume} {71}},\ \bibinfo {pages} {056501} (\bibinfo {year} {2008})}\BibitemShut {NoStop}%
\bibitem [{\citenamefont {Wang}\ \emph {et~al.}(2016)\citenamefont {Wang}, \citenamefont {Du}, \citenamefont {Liu}, \citenamefont {Hu}, \citenamefont {Zhang}, \citenamefont {Zhang}, \citenamefont {Owen}, \citenamefont {Lu}, \citenamefont {Gan}, \citenamefont {Sengupta}, \citenamefont {Kloc},\ and\ \citenamefont {Xiong}}]{Wang2016}%
  \BibitemOpen
  \bibfield  {author} {\bibinfo {author} {\bibfnamefont {X.}~\bibnamefont {Wang}}, \bibinfo {author} {\bibfnamefont {K.}~\bibnamefont {Du}}, \bibinfo {author} {\bibfnamefont {Y.~Y.~F.}\ \bibnamefont {Liu}}, \bibinfo {author} {\bibfnamefont {P.}~\bibnamefont {Hu}}, \bibinfo {author} {\bibfnamefont {J.}~\bibnamefont {Zhang}}, \bibinfo {author} {\bibfnamefont {Q.}~\bibnamefont {Zhang}}, \bibinfo {author} {\bibfnamefont {M.~H.~S.}\ \bibnamefont {Owen}}, \bibinfo {author} {\bibfnamefont {X.}~\bibnamefont {Lu}}, \bibinfo {author} {\bibfnamefont {C.~K.}\ \bibnamefont {Gan}}, \bibinfo {author} {\bibfnamefont {P.}~\bibnamefont {Sengupta}}, \bibinfo {author} {\bibfnamefont {C.}~\bibnamefont {Kloc}},\ and\ \bibinfo {author} {\bibfnamefont {Q.}~\bibnamefont {Xiong}},\ }\bibfield  {title} {\bibinfo {title} {Raman spectroscopy of atomically thin two-dimensional magnetic iron phosphorus trisulfide (\feps) crystals},\ }\href {https://doi.org/10.1088/2053-1583/3/3/031009} {\bibfield  {journal} {\bibinfo  {journal} {2D
  Materials}\ }\textbf {\bibinfo {volume} {3}},\ \bibinfo {pages} {031009} (\bibinfo {year} {2016})}\BibitemShut {NoStop}%
\bibitem [{\citenamefont {Zhang}\ \emph {et~al.}(2021{\natexlab{b}})\citenamefont {Zhang}, \citenamefont {Hwangbo}, \citenamefont {Wang}, \citenamefont {Jiang}, \citenamefont {Chu}, \citenamefont {Wen}, \citenamefont {Xiao},\ and\ \citenamefont {Xu}}]{Zhang2021-Giant_OLD}%
  \BibitemOpen
  \bibfield  {author} {\bibinfo {author} {\bibfnamefont {Q.}~\bibnamefont {Zhang}}, \bibinfo {author} {\bibfnamefont {K.}~\bibnamefont {Hwangbo}}, \bibinfo {author} {\bibfnamefont {C.}~\bibnamefont {Wang}}, \bibinfo {author} {\bibfnamefont {Q.}~\bibnamefont {Jiang}}, \bibinfo {author} {\bibfnamefont {J.-H.}\ \bibnamefont {Chu}}, \bibinfo {author} {\bibfnamefont {H.}~\bibnamefont {Wen}}, \bibinfo {author} {\bibfnamefont {D.}~\bibnamefont {Xiao}},\ and\ \bibinfo {author} {\bibfnamefont {X.}~\bibnamefont {Xu}},\ }\bibfield  {title} {\bibinfo {title} {Observation of giant optical linear dichroism in a zigzag antiferromagnet \feps},\ }\href {https://doi.org/10.1021/acs.nanolett.1c02188} {\bibfield  {journal} {\bibinfo  {journal} {Nano Letters}\ }\textbf {\bibinfo {volume} {21}},\ \bibinfo {pages} {6938} (\bibinfo {year} {2021}{\natexlab{b}})}\BibitemShut {NoStop}%
\bibitem [{\citenamefont {Markovin}\ and\ \citenamefont {Pisarev}(1979)}]{Markovin-JETP1979}%
  \BibitemOpen
  \bibfield  {author} {\bibinfo {author} {\bibfnamefont {P.}~\bibnamefont {Markovin}}\ and\ \bibinfo {author} {\bibfnamefont {R.}~\bibnamefont {Pisarev}},\ }\bibfield  {title} {\bibinfo {title} {Magnetic, thermal, and elastic refraction of light in the antiferromagnet {M}n{F}$_2$},\ }\href {http://jetp.ras.ru/cgi-bin/e/index/e/50/6/p1190?a=list} {\bibfield  {journal} {\bibinfo  {journal} {Sov. Phys. JETP}\ }\textbf {\bibinfo {volume} {50}},\ \bibinfo {pages} {1190} (\bibinfo {year} {1979})}\BibitemShut {NoStop}%
\bibitem [{\citenamefont {Saidl}\ \emph {et~al.}(2017)\citenamefont {Saidl}, \citenamefont {N{\v e}mec}, \citenamefont {Wadley}, \citenamefont {Hills}, \citenamefont {Campion}, \citenamefont {Nov{\'a}k}, \citenamefont {Edmonds}, \citenamefont {Maccherozzi}, \citenamefont {Dhesi}, \citenamefont {Gallagher}, \citenamefont {Troj{\'a}nek}, \citenamefont {Kune{\v s}}, \citenamefont {{\v Z}elezn{\'y}}, \citenamefont {Mal{\'y}},\ and\ \citenamefont {Jungwirth}}]{Saidl2017}%
  \BibitemOpen
  \bibfield  {author} {\bibinfo {author} {\bibfnamefont {V.}~\bibnamefont {Saidl}}, \bibinfo {author} {\bibfnamefont {P.}~\bibnamefont {N{\v e}mec}}, \bibinfo {author} {\bibfnamefont {P.}~\bibnamefont {Wadley}}, \bibinfo {author} {\bibfnamefont {V.}~\bibnamefont {Hills}}, \bibinfo {author} {\bibfnamefont {R.~P.}\ \bibnamefont {Campion}}, \bibinfo {author} {\bibfnamefont {V.}~\bibnamefont {Nov{\'a}k}}, \bibinfo {author} {\bibfnamefont {K.~W.}\ \bibnamefont {Edmonds}}, \bibinfo {author} {\bibfnamefont {F.}~\bibnamefont {Maccherozzi}}, \bibinfo {author} {\bibfnamefont {S.~S.}\ \bibnamefont {Dhesi}}, \bibinfo {author} {\bibfnamefont {B.~L.}\ \bibnamefont {Gallagher}}, \bibinfo {author} {\bibfnamefont {F.}~\bibnamefont {Troj{\'a}nek}}, \bibinfo {author} {\bibfnamefont {J.}~\bibnamefont {Kune{\v s}}}, \bibinfo {author} {\bibfnamefont {J.}~\bibnamefont {{\v Z}elezn{\'y}}}, \bibinfo {author} {\bibfnamefont {P.}~\bibnamefont {Mal{\'y}}},\ and\ \bibinfo {author} {\bibfnamefont {T.}~\bibnamefont {Jungwirth}},\
  }\bibfield  {title} {\bibinfo {title} {Optical determination of the {{N\'eel}} vector in a {{CuMnAs}} thin-film antiferromagnet},\ }\href {https://doi.org/10.1038/nphoton.2016.255} {\bibfield  {journal} {\bibinfo  {journal} {Nature Photonics}\ }\textbf {\bibinfo {volume} {11}},\ \bibinfo {pages} {91} (\bibinfo {year} {2017})}\BibitemShut {NoStop}%
\bibitem [{\citenamefont {Banda}(1986)}]{Banda1986}%
  \BibitemOpen
  \bibfield  {author} {\bibinfo {author} {\bibfnamefont {E.~J. K.~B.}\ \bibnamefont {Banda}},\ }\bibfield  {title} {\bibinfo {title} {Optical absorption of \nips in the near-infrared, visible and near-ultraviolet regions},\ }\href {https://doi.org/10.1088/0022-3719/19/36/023} {\bibfield  {journal} {\bibinfo  {journal} {Journal of Physics C: Solid State Physics}\ }\textbf {\bibinfo {volume} {19}},\ \bibinfo {pages} {7329} (\bibinfo {year} {1986})}\BibitemShut {NoStop}%
\bibitem [{\citenamefont {Joy}\ and\ \citenamefont {Vasudevan}(1992{\natexlab{b}})}]{Joy1992}%
  \BibitemOpen
  \bibfield  {author} {\bibinfo {author} {\bibfnamefont {P.~A.}\ \bibnamefont {Joy}}\ and\ \bibinfo {author} {\bibfnamefont {S.}~\bibnamefont {Vasudevan}},\ }\bibfield  {title} {\bibinfo {title} {Optical-absorption spectra of the layered transition-metal thiophosphates \mps ({M} ={M}n, {F}e, and {Ni})},\ }\href {https://doi.org/10.1103/PhysRevB.46.5134} {\bibfield  {journal} {\bibinfo  {journal} {Physical Review B}\ }\textbf {\bibinfo {volume} {46}},\ \bibinfo {pages} {5134} (\bibinfo {year} {1992}{\natexlab{b}})}\BibitemShut {NoStop}%
\bibitem [{\citenamefont {Demokritov}\ \emph {et~al.}(1987)\citenamefont {Demokritov}, \citenamefont {Kreines},\ and\ \citenamefont {Kudinov}}]{Demokritov-JETP1987}%
  \BibitemOpen
  \bibfield  {author} {\bibinfo {author} {\bibfnamefont {S.~O.}\ \bibnamefont {Demokritov}}, \bibinfo {author} {\bibfnamefont {N.~M.}\ \bibnamefont {Kreines}},\ and\ \bibinfo {author} {\bibfnamefont {V.~I.}\ \bibnamefont {Kudinov}},\ }\bibfield  {title} {\bibinfo {title} {Inelastic scattering of light in the antiferromagnet {EuTe}},\ }\href {http://jetp.ras.ru/cgi-bin/e/index/e/65/2/p389?a=list} {\bibfield  {journal} {\bibinfo  {journal} {Sov. Phys. JETP}\ }\textbf {\bibinfo {volume} {65}},\ \bibinfo {pages} {389} (\bibinfo {year} {1987})}\BibitemShut {NoStop}%
\bibitem [{\citenamefont {Landau}\ and\ \citenamefont {Lifshitz}(1980)}]{LANDAU1980446}%
  \BibitemOpen
  \bibfield  {author} {\bibinfo {author} {\bibfnamefont {L.}~\bibnamefont {Landau}}\ and\ \bibinfo {author} {\bibfnamefont {E.}~\bibnamefont {Lifshitz}},\ }\bibfield  {title} {\bibinfo {title} {Chapter {XIV} - phase transitions of the second kind and critical phenomena},\ }in\ \href@noop {} {\emph {\bibinfo {booktitle} {Statistical Physics}}},\ \bibinfo {editor} {edited by\ \bibinfo {editor} {\bibfnamefont {L.}~\bibnamefont {Landau}}\ and\ \bibinfo {editor} {\bibfnamefont {E.}~\bibnamefont {Lifshitz}}}\ (\bibinfo  {publisher} {Butterworth-Heinemann},\ \bibinfo {address} {Oxford},\ \bibinfo {year} {1980})\ \bibinfo {edition} {3rd}\ ed.,\ pp.\ \bibinfo {pages} {446--516}\BibitemShut {NoStop}%
\bibitem [{\citenamefont {Anisimov}\ \emph {et~al.}(1974)\citenamefont {Anisimov}, \citenamefont {Kapeliovich},\ and\ \citenamefont {Perelman}}]{anisimov1974electron}%
  \BibitemOpen
  \bibfield  {author} {\bibinfo {author} {\bibfnamefont {S.~I.}\ \bibnamefont {Anisimov}}, \bibinfo {author} {\bibfnamefont {B.~L.}\ \bibnamefont {Kapeliovich}},\ and\ \bibinfo {author} {\bibfnamefont {T.~L.}\ \bibnamefont {Perelman}},\ }\bibfield  {title} {\bibinfo {title} {Electron emission from metal surfaces exposed to ultrashort laser pulses},\ }\href {http://jetp.ras.ru/cgi-bin/e/index/e/39/2/p375?a=list} {\bibfield  {journal} {\bibinfo  {journal} {Sov. Phys. JETP}\ }\textbf {\bibinfo {volume} {39}},\ \bibinfo {pages} {375} (\bibinfo {year} {1974})}\BibitemShut {NoStop}%
\bibitem [{\citenamefont {Kirilyuk}\ \emph {et~al.}(2010)\citenamefont {Kirilyuk}, \citenamefont {Kimel},\ and\ \citenamefont {Rasing}}]{Kirilyuk-RMP2010}%
  \BibitemOpen
  \bibfield  {author} {\bibinfo {author} {\bibfnamefont {A.}~\bibnamefont {Kirilyuk}}, \bibinfo {author} {\bibfnamefont {A.~V.}\ \bibnamefont {Kimel}},\ and\ \bibinfo {author} {\bibfnamefont {T.}~\bibnamefont {Rasing}},\ }\bibfield  {title} {\bibinfo {title} {Ultrafast optical manipulation of magnetic order},\ }\href {https://doi.org/10.1103/RevModPhys.82.2731} {\bibfield  {journal} {\bibinfo  {journal} {Rev. Mod. Phys.}\ }\textbf {\bibinfo {volume} {82}},\ \bibinfo {pages} {2731} (\bibinfo {year} {2010})}\BibitemShut {NoStop}%
\bibitem [{\citenamefont {Debye}(1912)}]{debye_zur_1912}%
  \BibitemOpen
  \bibfield  {author} {\bibinfo {author} {\bibfnamefont {P.}~\bibnamefont {Debye}},\ }\bibfield  {title} {\bibinfo {title} {Zur {Theorie} der spezifischen {W\"armen}},\ }\href {https://doi.org/10.1002/andp.19123441404} {\bibfield  {journal} {\bibinfo  {journal} {Annalen der Physik}\ }\textbf {\bibinfo {volume} {344}},\ \bibinfo {pages} {789} (\bibinfo {year} {1912})}\BibitemShut {NoStop}%
\bibitem [{\citenamefont {Walowski}(2012)}]{walowski2012PhD}%
  \BibitemOpen
  \bibfield  {author} {\bibinfo {author} {\bibfnamefont {J.}~\bibnamefont {Walowski}},\ }\emph {\bibinfo {title} {Physics of laser heated ferromagnets: Ultrafast demagnetization and magneto-{S}eebeck effect}},\ \href {https://doi.org/10.53846/goediss-2944} {Ph.D. thesis},\ \bibinfo  {school} {Georg-August-University G\"ottingen} (\bibinfo {year} {2012})\BibitemShut {NoStop}%
\bibitem [{\citenamefont {Jain}\ \emph {et~al.}(2013)\citenamefont {Jain}, \citenamefont {Ong}, \citenamefont {Hautier}, \citenamefont {Chen}, \citenamefont {Richards}, \citenamefont {Dacek}, \citenamefont {Cholia}, \citenamefont {Gunter}, \citenamefont {Skinner}, \citenamefont {Ceder},\ and\ \citenamefont {Persson}}]{Jain2013_MP}%
  \BibitemOpen
  \bibfield  {author} {\bibinfo {author} {\bibfnamefont {A.}~\bibnamefont {Jain}}, \bibinfo {author} {\bibfnamefont {S.~P.}\ \bibnamefont {Ong}}, \bibinfo {author} {\bibfnamefont {G.}~\bibnamefont {Hautier}}, \bibinfo {author} {\bibfnamefont {W.}~\bibnamefont {Chen}}, \bibinfo {author} {\bibfnamefont {W.~D.}\ \bibnamefont {Richards}}, \bibinfo {author} {\bibfnamefont {S.}~\bibnamefont {Dacek}}, \bibinfo {author} {\bibfnamefont {S.}~\bibnamefont {Cholia}}, \bibinfo {author} {\bibfnamefont {D.}~\bibnamefont {Gunter}}, \bibinfo {author} {\bibfnamefont {D.}~\bibnamefont {Skinner}}, \bibinfo {author} {\bibfnamefont {G.}~\bibnamefont {Ceder}},\ and\ \bibinfo {author} {\bibfnamefont {K.~A.}\ \bibnamefont {Persson}},\ }\bibfield  {title} {\bibinfo {title} {Commentary: The materials project: A materials genome approach to accelerating materials innovation},\ }\href {https://doi.org/10.1063/1.4812323} {\bibfield  {journal} {\bibinfo  {journal} {APL Materials}\ }\textbf {\bibinfo {volume} {1}},\ \bibinfo {pages}
  {011002} (\bibinfo {year} {2013})}\BibitemShut {NoStop}%
\bibitem [{\citenamefont {Piacentini}\ \emph {et~al.}(1982)\citenamefont {Piacentini}, \citenamefont {Khumalo}, \citenamefont {Olson}, \citenamefont {Anderegg},\ and\ \citenamefont {Lynch}}]{Piacentini1982}%
  \BibitemOpen
  \bibfield  {author} {\bibinfo {author} {\bibfnamefont {M.}~\bibnamefont {Piacentini}}, \bibinfo {author} {\bibfnamefont {F.~S.}\ \bibnamefont {Khumalo}}, \bibinfo {author} {\bibfnamefont {C.~G.}\ \bibnamefont {Olson}}, \bibinfo {author} {\bibfnamefont {J.~W.}\ \bibnamefont {Anderegg}},\ and\ \bibinfo {author} {\bibfnamefont {D.~W.}\ \bibnamefont {Lynch}},\ }\bibfield  {title} {\bibinfo {title} {Optical transitions, xps, and electronic states in \nips},\ }\href {https://doi.org/10.1016/0301-0104(82)85205-1} {\bibfield  {journal} {\bibinfo  {journal} {Chemical Physics}\ }\textbf {\bibinfo {volume} {65}},\ \bibinfo {pages} {289} (\bibinfo {year} {1982})}\BibitemShut {NoStop}%
\bibitem [{\citenamefont {Zhang}\ \emph {et~al.}(2022)\citenamefont {Zhang}, \citenamefont {Ni}, \citenamefont {Stevens}, \citenamefont {Bai}, \citenamefont {Peiris}, \citenamefont {Hendrickson}, \citenamefont {Wu},\ and\ \citenamefont {Jariwala}}]{Zhang2022}%
  \BibitemOpen
  \bibfield  {author} {\bibinfo {author} {\bibfnamefont {H.}~\bibnamefont {Zhang}}, \bibinfo {author} {\bibfnamefont {Z.}~\bibnamefont {Ni}}, \bibinfo {author} {\bibfnamefont {C.~E.}\ \bibnamefont {Stevens}}, \bibinfo {author} {\bibfnamefont {A.}~\bibnamefont {Bai}}, \bibinfo {author} {\bibfnamefont {F.}~\bibnamefont {Peiris}}, \bibinfo {author} {\bibfnamefont {J.~R.}\ \bibnamefont {Hendrickson}}, \bibinfo {author} {\bibfnamefont {L.}~\bibnamefont {Wu}},\ and\ \bibinfo {author} {\bibfnamefont {D.}~\bibnamefont {Jariwala}},\ }\bibfield  {title} {\bibinfo {title} {Cavity-enhanced linear dichroism in a van der {W}aals antiferromagnet},\ }\href {https://doi.org/10.1038/s41566-022-00970-8} {\bibfield  {journal} {\bibinfo  {journal} {Nature Photonics}\ }\textbf {\bibinfo {volume} {16}},\ \bibinfo {pages} {311} (\bibinfo {year} {2022})}\BibitemShut {NoStop}%
\bibitem [{\citenamefont {Belvin}\ \emph {et~al.}(2021)\citenamefont {Belvin}, \citenamefont {Baldini}, \citenamefont {Ozel}, \citenamefont {Mao}, \citenamefont {Po}, \citenamefont {Allington}, \citenamefont {Son}, \citenamefont {Kim}, \citenamefont {Kim}, \citenamefont {Hwang}, \citenamefont {Kim}, \citenamefont {Park}, \citenamefont {Senthil},\ and\ \citenamefont {Gedik}}]{Belvin2021}%
  \BibitemOpen
  \bibfield  {author} {\bibinfo {author} {\bibfnamefont {C.~A.}\ \bibnamefont {Belvin}}, \bibinfo {author} {\bibfnamefont {E.}~\bibnamefont {Baldini}}, \bibinfo {author} {\bibfnamefont {I.~O.}\ \bibnamefont {Ozel}}, \bibinfo {author} {\bibfnamefont {D.}~\bibnamefont {Mao}}, \bibinfo {author} {\bibfnamefont {H.~C.}\ \bibnamefont {Po}}, \bibinfo {author} {\bibfnamefont {C.~J.}\ \bibnamefont {Allington}}, \bibinfo {author} {\bibfnamefont {S.}~\bibnamefont {Son}}, \bibinfo {author} {\bibfnamefont {B.~H.}\ \bibnamefont {Kim}}, \bibinfo {author} {\bibfnamefont {J.}~\bibnamefont {Kim}}, \bibinfo {author} {\bibfnamefont {I.}~\bibnamefont {Hwang}}, \bibinfo {author} {\bibfnamefont {J.~H.}\ \bibnamefont {Kim}}, \bibinfo {author} {\bibfnamefont {J.~G.}\ \bibnamefont {Park}}, \bibinfo {author} {\bibfnamefont {T.}~\bibnamefont {Senthil}},\ and\ \bibinfo {author} {\bibfnamefont {N.}~\bibnamefont {Gedik}},\ }\bibfield  {title} {\bibinfo {title} {Exciton-driven antiferromagnetic metal in a correlated van der {W}aals
  insulator},\ }\bibfield  {journal} {\bibinfo  {journal} {Nature Communications}\ }\textbf {\bibinfo {volume} {12}},\ \href {https://doi.org/10.1038/s41467-021-25164-8} {10.1038/s41467-021-25164-8} (\bibinfo {year} {2021})\BibitemShut {NoStop}%
\bibitem [{\citenamefont {Budniak}\ \emph {et~al.}(2022)\citenamefont {Budniak}, \citenamefont {Zelewski}, \citenamefont {Birowska}, \citenamefont {Woźniak}, \citenamefont {Bendikov}, \citenamefont {Kauffmann}, \citenamefont {Amouyal}, \citenamefont {Kudrawiec},\ and\ \citenamefont {Lifshitz}}]{Budniak2022}%
  \BibitemOpen
  \bibfield  {author} {\bibinfo {author} {\bibfnamefont {A.~K.}\ \bibnamefont {Budniak}}, \bibinfo {author} {\bibfnamefont {S.~J.}\ \bibnamefont {Zelewski}}, \bibinfo {author} {\bibfnamefont {M.}~\bibnamefont {Birowska}}, \bibinfo {author} {\bibfnamefont {T.}~\bibnamefont {Woźniak}}, \bibinfo {author} {\bibfnamefont {T.}~\bibnamefont {Bendikov}}, \bibinfo {author} {\bibfnamefont {Y.}~\bibnamefont {Kauffmann}}, \bibinfo {author} {\bibfnamefont {Y.}~\bibnamefont {Amouyal}}, \bibinfo {author} {\bibfnamefont {R.}~\bibnamefont {Kudrawiec}},\ and\ \bibinfo {author} {\bibfnamefont {E.}~\bibnamefont {Lifshitz}},\ }\bibfield  {title} {\bibinfo {title} {Spectroscopy and structural investigation of iron phosphorus trisulfide—\feps},\ }\bibfield  {journal} {\bibinfo  {journal} {Advanced Optical Materials}\ }\textbf {\bibinfo {volume} {10}},\ \href {https://doi.org/10.1002/adom.202102489} {10.1002/adom.202102489} (\bibinfo {year} {2022})\BibitemShut {NoStop}%
\bibitem [{\citenamefont {Jos\'e}\ \emph {et~al.}(1977)\citenamefont {Jos\'e}, \citenamefont {Kadanoff}, \citenamefont {Kirkpatrick},\ and\ \citenamefont {Nelson}}]{Jose-PRB1977}%
  \BibitemOpen
  \bibfield  {author} {\bibinfo {author} {\bibfnamefont {J.~V.}\ \bibnamefont {Jos\'e}}, \bibinfo {author} {\bibfnamefont {L.~P.}\ \bibnamefont {Kadanoff}}, \bibinfo {author} {\bibfnamefont {S.}~\bibnamefont {Kirkpatrick}},\ and\ \bibinfo {author} {\bibfnamefont {D.~R.}\ \bibnamefont {Nelson}},\ }\bibfield  {title} {\bibinfo {title} {Renormalization, vortices, and symmetry-breaking perturbations in the two-dimensional planar model},\ }\href {https://doi.org/10.1103/PhysRevB.16.1217} {\bibfield  {journal} {\bibinfo  {journal} {Phys. Rev. B}\ }\textbf {\bibinfo {volume} {16}},\ \bibinfo {pages} {1217} (\bibinfo {year} {1977})}\BibitemShut {NoStop}%
\bibitem [{\citenamefont {Yao-Dong}\ \emph {et~al.}(2004)\citenamefont {Yao-Dong}, \citenamefont {Lin}, \citenamefont {Ya-Xin}, \citenamefont {Yun}, \citenamefont {Hong-Bo},\ and\ \citenamefont {Yuan-Fu}}]{Yao-Dong2004}%
  \BibitemOpen
  \bibfield  {author} {\bibinfo {author} {\bibfnamefont {D.}~\bibnamefont {Yao-Dong}}, \bibinfo {author} {\bibfnamefont {W.}~\bibnamefont {Lin}}, \bibinfo {author} {\bibfnamefont {Y.}~\bibnamefont {Ya-Xin}}, \bibinfo {author} {\bibfnamefont {H.}~\bibnamefont {Yun}}, \bibinfo {author} {\bibfnamefont {H.}~\bibnamefont {Hong-Bo}},\ and\ \bibinfo {author} {\bibfnamefont {X.}~\bibnamefont {Yuan-Fu}},\ }\bibfield  {title} {\bibinfo {title} {A {M}\"ossbauer study of the magnetic coupling in iron phosphorous trisulfides},\ }\href {https://doi.org/10.1088/1009-1963/13/10/013} {\bibfield  {journal} {\bibinfo  {journal} {Chinese Physics}\ }\textbf {\bibinfo {volume} {13}},\ \bibinfo {pages} {1652} (\bibinfo {year} {2004})}\BibitemShut {NoStop}%
\bibitem [{\citenamefont {Takano}\ \emph {et~al.}(2004)\citenamefont {Takano}, \citenamefont {Arai}, \citenamefont {Arai}, \citenamefont {Takahashi}, \citenamefont {Takase},\ and\ \citenamefont {Sekizawa}}]{TAKANO2004}%
  \BibitemOpen
  \bibfield  {author} {\bibinfo {author} {\bibfnamefont {Y.}~\bibnamefont {Takano}}, \bibinfo {author} {\bibfnamefont {N.}~\bibnamefont {Arai}}, \bibinfo {author} {\bibfnamefont {A.}~\bibnamefont {Arai}}, \bibinfo {author} {\bibfnamefont {Y.}~\bibnamefont {Takahashi}}, \bibinfo {author} {\bibfnamefont {K.}~\bibnamefont {Takase}},\ and\ \bibinfo {author} {\bibfnamefont {K.}~\bibnamefont {Sekizawa}},\ }\bibfield  {title} {\bibinfo {title} {Magnetic properties and specific heat of {MPS}$_3$ ({M}={M}n, {F}e, {Z}n)},\ }\href {https://doi.org/10.1016/j.jmmm.2003.12.621} {\bibfield  {journal} {\bibinfo  {journal} {Journal of Magnetism and Magnetic Materials}\ }\textbf {\bibinfo {volume} {272-276}},\ \bibinfo {pages} {E593} (\bibinfo {year} {2004})}\BibitemShut {NoStop}%
\bibitem [{\citenamefont {Minnich}(2016)}]{Minnich2016}%
  \BibitemOpen
  \bibfield  {author} {\bibinfo {author} {\bibfnamefont {A.~J.}\ \bibnamefont {Minnich}},\ }\bibfield  {title} {\bibinfo {title} {Exploring the extremes of heat conduction in anisotropic materials},\ }\href {https://doi.org/10.1080/15567265.2016.1170080} {\bibfield  {journal} {\bibinfo  {journal} {Nanoscale and Microscale Thermophysical Engineering}\ }\textbf {\bibinfo {volume} {20}},\ \bibinfo {pages} {1} (\bibinfo {year} {2016})}\BibitemShut {NoStop}%
\bibitem [{\citenamefont {Matthiesen}\ \emph {et~al.}(2023)\citenamefont {Matthiesen}, \citenamefont {Hortensius}, \citenamefont {Ma{\~n}as-Valero}, \citenamefont {Kapon}, \citenamefont {Dumcenco}, \citenamefont {Giannini}, \citenamefont {\ifmmode \check{S}\else \v{S}\fi{}i\ifmmode~\check{s}\else \v{s}\fi{}kins}, \citenamefont {Ivanov}, \citenamefont {van~der Zant}, \citenamefont {Coronado}, \citenamefont {Kuzmenko}, \citenamefont {Afanasiev},\ and\ \citenamefont {Caviglia}}]{Matthiesen2023}%
  \BibitemOpen
  \bibfield  {author} {\bibinfo {author} {\bibfnamefont {M.}~\bibnamefont {Matthiesen}}, \bibinfo {author} {\bibfnamefont {J.~R.}\ \bibnamefont {Hortensius}}, \bibinfo {author} {\bibfnamefont {S.}~\bibnamefont {Ma{\~n}as-Valero}}, \bibinfo {author} {\bibfnamefont {I.}~\bibnamefont {Kapon}}, \bibinfo {author} {\bibfnamefont {D.}~\bibnamefont {Dumcenco}}, \bibinfo {author} {\bibfnamefont {E.}~\bibnamefont {Giannini}}, \bibinfo {author} {\bibfnamefont {M.}~\bibnamefont {\ifmmode \check{S}\else \v{S}\fi{}i\ifmmode~\check{s}\else \v{s}\fi{}kins}}, \bibinfo {author} {\bibfnamefont {B.~A.}\ \bibnamefont {Ivanov}}, \bibinfo {author} {\bibfnamefont {H.~S.~J.}\ \bibnamefont {van~der Zant}}, \bibinfo {author} {\bibfnamefont {E.}~\bibnamefont {Coronado}}, \bibinfo {author} {\bibfnamefont {A.~B.}\ \bibnamefont {Kuzmenko}}, \bibinfo {author} {\bibfnamefont {D.}~\bibnamefont {Afanasiev}},\ and\ \bibinfo {author} {\bibfnamefont {A.~D.}\ \bibnamefont {Caviglia}},\ }\bibfield  {title} {\bibinfo {title} {Controlling magnetism
  with light in a zero orbital angular momentum antiferromagnet},\ }\href {https://doi.org/10.1103/PhysRevLett.130.076702} {\bibfield  {journal} {\bibinfo  {journal} {Phys. Rev. Lett.}\ }\textbf {\bibinfo {volume} {130}},\ \bibinfo {pages} {076702} (\bibinfo {year} {2023})}\BibitemShut {NoStop}%
\bibitem [{\citenamefont {Wildes}\ \emph {et~al.}(2017)\citenamefont {Wildes}, \citenamefont {Simonet}, \citenamefont {Ressouche}, \citenamefont {Ballou},\ and\ \citenamefont {McIntyre}}]{Wildes_JPCM2017}%
  \BibitemOpen
  \bibfield  {author} {\bibinfo {author} {\bibfnamefont {A.~R.}\ \bibnamefont {Wildes}}, \bibinfo {author} {\bibfnamefont {V.}~\bibnamefont {Simonet}}, \bibinfo {author} {\bibfnamefont {E.}~\bibnamefont {Ressouche}}, \bibinfo {author} {\bibfnamefont {R.}~\bibnamefont {Ballou}},\ and\ \bibinfo {author} {\bibfnamefont {G.~J.}\ \bibnamefont {McIntyre}},\ }\bibfield  {title} {\bibinfo {title} {The magnetic properties and structure of the quasi-two-dimensional antiferromagnet \cops},\ }\href {https://doi.org/10.1088/1361-648X/aa8a43} {\bibfield  {journal} {\bibinfo  {journal} {Journal of Physics: Condensed Matter}\ }\textbf {\bibinfo {volume} {29}},\ \bibinfo {pages} {455801} (\bibinfo {year} {2017})}\BibitemShut {NoStop}%
\bibitem [{\citenamefont {Koopmans}\ \emph {et~al.}(2000)\citenamefont {Koopmans}, \citenamefont {van Kampen}, \citenamefont {Kohlhepp},\ and\ \citenamefont {de~Jonge}}]{Koopmans-PRL2000}%
  \BibitemOpen
  \bibfield  {author} {\bibinfo {author} {\bibfnamefont {B.}~\bibnamefont {Koopmans}}, \bibinfo {author} {\bibfnamefont {M.}~\bibnamefont {van Kampen}}, \bibinfo {author} {\bibfnamefont {J.~T.}\ \bibnamefont {Kohlhepp}},\ and\ \bibinfo {author} {\bibfnamefont {W.~J.~M.}\ \bibnamefont {de~Jonge}},\ }\bibfield  {title} {\bibinfo {title} {Ultrafast magneto-optics in nickel: Magnetism or optics?},\ }\href {https://doi.org/10.1103/PhysRevLett.85.844} {\bibfield  {journal} {\bibinfo  {journal} {Phys. Rev. Lett.}\ }\textbf {\bibinfo {volume} {85}},\ \bibinfo {pages} {844} (\bibinfo {year} {2000})}\BibitemShut {NoStop}%
\end{thebibliography}%

\end{document}